\documentclass[11pt]{article}
\usepackage{amsmath,epsfig,amscd,amssymb}
\usepackage{latexsym}
\usepackage{graphicx}

\renewcommand{\thesection}{\arabic{section}.}
\renewcommand{\thesubsection}{\arabic{section}.\arabic{subsection}.}
\renewcommand{\theequation}{\arabic{section}.\arabic{equation}}
\renewcommand{\thesubsubsection}
{\arabic{section}.\arabic{subsection}.\arabic{subsubsection}.}

\begin{document}


{\Large \bf Assessment and Propagation of Input Uncertainty in
Tree-based Option Pricing Models}
\vspace{-0.1in}

\small{
\begin{tabbing}
\hspace{.8in} \= Henryk Gzyl \hspace{.9in} \= German Molina
\hspace{.9in} \= Enrique ter Horst \\

\> IESA  \> Vega Capital Services Ltd.  \> IESA \\
\end{tabbing}
}


\begin{tabbing}
\hspace{2.2in} \= \textbf{Abstract}
\end{tabbing}
This paper aims to provide a practical example
on the assessment and propagation of input uncertainty for option
pricing when using tree-based methods. Input uncertainty is
propagated into output uncertainty, reflecting that
option prices are as unknown as the inputs they are based on.
Option pricing formulas are tools whose validity is conditional
not only on how close the model represents reality, but also on
the quality of the inputs they use, and those inputs are usually
not observable. We provide three alternative frameworks to calibrate option pricing tree models, propagating
parameter uncertainty into the resulting option prices. We finally compare our
methods with classical calibration-based results assuming that there is no options market established. These methods can
be applied to pricing of instruments for which there is not an
options market, as well as a methodological tool to account for
parameter and model uncertainty in theoretical option pricing.

\begin{quotation}

\end{quotation}
\section*{Acknowledgements}
We wish to thank Lars Stentoft and The Centre for Analytical
Finance, University of Aarhus Denmark for providing us with the
S$\&$P 500 option data; and Sabatino Constanzo, Samuel Malone,
Miguel Mayo, Abel Rodriguez, and Loren Trigo for providing us
helpful comments.

\vspace{-.15in}

\section{Introduction}
\setcounter{equation}{0}
\subsection{Option pricing dependencies}
Option pricing has become a major topic in quantitative finance.
Pricing models and algorithms, ranging from the now-basic Black \&
Scholes (1973) to more complex Partial Integro Differential
Equations (PIDE) (Cont et al., 2004) and tree-based methods (Cox
et al., (1979), Gustafsson et al., (2002)) are being proposed
continuously in what has become a very extensive
literature in mathematical finance. \\

All these option pricing models rely on some set of inputs, obtained
by either estimation (Polson et al., (2003)) or calibration (Cont et
al., (2004)) to implied market values, with most of them relying on
no-arbitrage arguments. Garcia et al., (2003) provides a good econometric review. \\

Once a set of input values is determined, they are passed to complex
mathematical functions, whenever closed-form solutions are
available, or computational algorithms to determine anything from
plain vanilla to very exotic option prices. This common framework
provides, in most cases, a unique solution, a unique option price
that practitioners consider the theoretical, risk-neutral value of
the option, and around which the market players will add the risk
premium and spread. However, throughout all this process, we must
not forget that the quality of such output relies extremely on the
quality of the inputs used.

In most cases, a key input, expressed one way or another, is the
volatility of the underlying asset (or combination of assets) that
defines the option. The realized volatility literature (Andersen
et al., 2003) has brought us closer to making this parameter
locally observable at higher frequencies. Some problems still
persist such as the lack of data, and the existence of other
parameters (nuisance parameters), which render the task of making
accurate assessments of our uncertainty about inputs for option
pricing models more difficult. In practice, it is common to use
the most likely input values or calibrated input values to price
options, and focus the efforts on good modelling of parameter dynamics.

During the last few years, there have been many advances in the
modelling of the underlying (stochastic volatility (Jacquier et al.,
(1994)), jump-diffusion models (Duffie et al., (2000))) and/or
modelling jointly the underlying and the observed options movements
(Eraker et al., 2003, Barndorff-Nielsen and Shephard (2001)).
However, no matter how accurate our models become to estimate
unobservable inputs, a proper accounting of their flaws and pitfalls
as well as assessment of input uncertainty is as important to option
pricing as the quality of the pricing model itself. Propagation of
input uncertainty through complex mathematical model output
uncertainty has been explored in other fields, like traffic
engineering (Molina et al., 2005) or climatology (Coles and Pericchi
(2003)).

The focus of this paper is the assessment and propagation of input
uncertainty, and its effects on option pricing through the
computation of a posterior distribution of the option prices,
conditional on the observed data, that we use to integrate out the
parameter's uncertainty from the option functional. That is, we look for
option prices that are unconditional from the parameters they rely on.
We illustrate
this idea through an alternative way to calibrate a tree-based model.
By using the historical returns of the underlying, and
looking at the up/down returns with respect to the risk-free rate,
but by using a related statistical parametric approach, we not only from the
sample variation of the up/down returns, but also from the scale
and location parameters from the return probability generating
process as well. As time passes by, we are able to observe more
returns, and adjust our knowledge about the parameters (assuming that they are constant).
We undertake a bayesian approach to the modelling and the estimation of the underlying (Hastings (1970)).
Parameter posterior distributions lead us to posterior
probabilities on the trees that we use afterwards to price
options. However, our goal is not to provide a better pricing
model but a new method to estimate model parameters and propagate
parameter uncertainty through the use of Bayesian methods,
illustrating how uncertainty of inputs can be reflected into
uncertainty of outputs. Therefore, we confine our application to the
tree-based classic approach of Cox et al., (1979), and build a statistical
model that accounts for input uncertainty, and propagate it into
option price uncertainty under the framework set by that model.//
In sections two and three we motivate
our approach to assessment and propagation of input uncertainty
into pricing model outputs together with a basic
decision-theoretic argument, and show that most likely parameter
values do not necessarily lead to optimal option price reference
choices. In section four we propose a statistical method for the
estimation of such input uncertainty and construct the link
between the statistical model, input uncertainty and the
Cox-Ross-Rubinstein (CRR) model. Section five contains an
application of this method for the S\&P500 as well as a comparison
with bootstrap-based calibrations of the tree model, together with
potential further applications in section 6. Finally, section 7
concludes. We will consider throughout the paper the situation where there
are no options markets in place to produce a better calibration, and option prices
must be developed solely from the information contained in the underlying instrument.

\section{Assessment and Propagation of input uncertainty}
\subsection{Motivation}
Why should parameter uncertainty matter? Does it really make a
difference for option pricing? After all, if we fit a model to the
most likely values through classical maximum likelihood methods, we
should be as close as we can to the "true" option theoretical,
risk-neutral value by the invariance principle of the MLE estimator
(assuming one-to-one relationships between input and output).
Furthermore, most practitioners want a simple, unique answer as to
what the market price should be.

There are four major reasons why we might not know the true value
of an option: First, our model might incorrectly represent the
dynamics of the underlying (Cont, R. (2006)) leading to incorrect
and biased option prices. We choose to ignore model uncertainty in
this paper (understanding models as different pricing tools, and not as different
trees), although it should be accounted for through the use of
model averaging approaches (Hoeting et al., (1999), Cont, R.
(2006)). Second, even if our option pricing model was correct, the
option's payoff is random. However, we can price the latter under
risk-neutral assumptions. Third, which is the focus of our paper,
even if the model was correct, the model parameters are not known,
and different combinations of inputs lead to different values of the
option. Finally, when we have more than one parameter defining the trees, the one-to-one relationship
between parameters and option prices breaks down.

The true value of options is not in terms of actual discounted
payoffs, but in terms of expected ones. For example, the value of a
digital option is either zero or some fixed amount, depending on the
path of the underlying until maturity, but what we try to model is
the expected value of that option as of today. The theoretical value
of the option will be considered known if we were to know the exact
value of the inputs that drive the underlying, but not the realized
path of the underlying. When pricing options, the most we can aspire
to learn from the dynamics of the underlying are the model
parameters, as the path will remain stochastic. That is why we
define "value" of an option at a given period only in terms of the
risk-neutral valuation based on the true, unknown inputs. Even if
the model accurately represents the dynamics of the underlying, the
model parameters are still unknown. This is especially true as we
construct more complicated models, where parameters could not only
be dynamic but stochastic as well, and even an infinite amount of
data would not suffice to learn about their future values.

It is common for practitioners to price options by using the mode of
the inputs' probability distributions (eg most likely value of the
volatility). A first glitch comes when mapping a multi-dimensional
input parameter vector into a 1-dimensional output, as the most
likely value of the input does not necessarily lead to the most
likely value of the output. This is a strong argument in favor of
finding approaches that reflect actual parameter uncertainty. All we
want to know is an option´s most likely value? Perhaps its value
based on the most likely input? What about its expected value? If we
obtain the option price's probability distribution, we can extract
much more information, including but not restricted to all those.

\subsection{How to propagate input uncertainty}
Calibrating a model with the most likely input value does not lead
to the "true" option value, nor to the expected/optimal option
value. The most likely value of the parameter has probability zero
of being the true value in continuous parameter spaces. This is
quite important, since option prices are asymmetric with respect to
most of their inputs, making the effect on the option price of a small parameter error
in either direction very different.

We illustrate this idea with a simple example. Suppose that the
volatility for an underlying can take only three possible values:
$\hat{v}-a$ with probability 30\%, $\hat{v}$ with probability 40\% ,
and $\hat{v}+a$ with probability 30\%. $\hat{v}$ is not only the
most likely value, but also the expected value. Without loss of
generality, we assume that this is the only parameter needed to
price an option. The option theoretical, risk-neutral price with
volatility $\hat{v}$ is equal to $P(\hat{v})$. However, if the true
volatility were to be $\hat{v}-a$, $\hat{v}$ or $\hat{v}+a$
respectively, the option price would be equal to
$P(\hat{v}-a)<P(\hat{v})<P(\hat{v}+a)$. Due to the asymmetric effect
of the inputs over the outputs, we know that
$P(\hat{v})-P(\hat{v}-a)\neq{}P(\hat{v}+a)-P(\hat{v})$. Therefore,
although $\hat{v}$ is both the most likely and the expected value
for the input, $P(\hat{v})$ is not the expected value of the output.
To what extent should we use $\hat{v}$ to price this option? The
expected option price is not $P(\hat{v})$, but a larger value, even
under this symmetric and relatively nice distribution of the input
$v$. Considering that 60\% of the time we will be wrong by choosing
the most likely value for the input, we need to consider not only
one possible value, but all possible values of the inputs, together
with their probability distributions when assessing the possible
values for the output. The question now becomes how to propagate the
uncertainty about $v$ into a final option price for more general
settings.

This argument can be formalized more properly. Suppose that,
instead of three possible values, the parameter $v$ has a
probability distribution of possible values. Therefore, for each
value of $v$, we have a possible option price $P(v)$, with
probability $\pi(v)$. This would represent the uncertainty about
the output $P$ as a function of the input´s uncertainty $v$.

Even if we wanted a single option value as an answer, we can still
do so by integrating out the option price with respect to the
distribution $\pi(v)$ yielding $E_{\pi}(P) =
\int{P(v)\pi(v)dv}\neq{P(\int{v\pi(v)dv})}\neq{}P(\hat{v})$, where
the first expression is the expected option value, the second is the
price under the expected parameter value, and the third is the
price under the most likely parameter value. Our focus is on the
first expression. This expected option value propagates the input's
uncertainty when passing it to a final option price that is not
dependent on a single volatility value $\hat{v}$ or $E_{\pi}(v)$,
but rather on the overall features of the input's probability
distribution $\pi(v)$ and their effect on the output pricing
formula. The resulting option price is no-longer conditional on the
volatility, but marginalized over it. The estimation problem has not
disappeared, but has been transformed. We no longer focus on a
single estimate $\hat{v}$, but rather on the assessment of the
input´s uncertainty through $\pi(v)$. In other words, our focus
changes from how much we know (most likely value) to how much we do
not know (probability distribution). In general a unique option
value deprives us from a full assessment of the uncertainty in
option valuation. It is not the same to know that the option price
is with 99\% probability between 40 and 42 than if it is between 37
and 45, even if in both cases the expected and most likely values
happened to be the same. Market players will act differently on
those two cases. This happens when the option value is asymmetric,
as we will show later in the paper.

We can naturally generalize the formulae above to several
parameters, making the expected option price $P$ a mere integral
over the probability distributions of the possible
parameters/inputs. $E_{\pi}(P) = \int{P(\xi)\pi(\xi)d\xi}$, where
$\xi=(d,u)$ is a bi-dimensional vector that calibrates a tree to
model the price of an underlying, where $u$ and $d$ are the upward and downward returns
respectively. We must here assess the (joint)
probability distribution for all the parameters $\pi(\xi)$ as
parameter correlations influence inferential results. Assessing
parameter values is of special difficulty when either limited data
is available or the dynamics of the underlying are difficult to
model, leading to inferential problems.
\subsection{Uncertainty estimation as a feature in pricing tools}
\label{uncertainty}

We outlined how the importance of uncertainty estimation impacts
the option value through a probability distribution of the inputs
$\xi$. In practice, such a probability distribution must be
estimated/updated using available data for the underlying, more so when options
markets do not yet exist.

Our approach bears on model uncertainty. Following Cont (2006), if
we regard the data on the prices of the underlying as prior data
at $t = 0$, the posterior probabilities describe the model
uncertainty regarding the option pricing model through the
posterior probability distribution of $\xi$, or the model
misspecification. Bayesian model averaging as described in Hoeting
et al., (1999), Cont (2006) is thus a way to incorporate model
uncertainty .

We take a Bayesian approach to the statistical estimation of the
CRR model in this paper. This has several advantages. First, it
allows for prior information to be naturally included whenever
available. This is especially useful in situations where the data
is scarce, distributions vary over time or we cannot rely on
asymptotics for calibration. Second, it provides a natural way to
account for uncertainty in the parameters after we observe the
underlying's historical return series, and the necessary dynamic
updating as new information becomes available. The posterior
distribution of parameters given the data fits conceptually well
into this framework, allowing to sequentially update and learn
about the parameters $\xi$ of the tree-generating process.

Let $X_t$ be the data available regarding the observable
$\xi$ to make inference about its parameter vector $\theta$
at time $t$, linked through the likelihood function
$f(X_t|\theta)$. Without loss of generality, assume that the
inputs are not time-dependent. Given the data and any additional
information regarding the parameter vector $\theta$, prior to
collecting that data, $\pi(\theta)$, we can update that
information with the data to obtain the posterior distribution for
$\theta$: $\pi(\theta|X_t)\propto{f(X_t|\theta)\pi(\theta)}$. One
can then use this probability distribution to propagate our
uncertainty about $\xi$ through its posterior distribution once
$\theta$ is integrated out in the following way:
$$\pi(\xi|X_t)=\int_{\Theta}f(\xi|\theta)\pi(\theta|X_t)d\theta$$

The posterior distribution for $\xi$ given the data will in turn
propage the uncertainty into the option prices, in the following
way:
$$E_{\pi}(P) = \int{P(\xi)\pi(\xi|X_t)d\xi}.$$
\noindent The likelihood function $f(X_t|\theta)$ (under the
physical measure) becomes the main tool to obtain uncertainty
estimates about the parameters $\theta$, and, consequently, about
the option prices $P$. The likelihood function must be consistent
with the option pricing tools, as the parameters must have the
same meaning under both, or allow some mapping from the physical
measure $\mathbb{P}$ to the risk neutral measure $\mathbb{Q}$.
Therefore, the function $f$ is model specific, and must be
constructed accordingly, not only to properly reflect the dynamics
of the underlying, but also its relationship with the inputs as
defined in the option pricing model.

\section{Decision-Theoretic justification}\label{utilities}
\subsection{Motivation}
We assume that our pricing model perfectly characterizes the
dynamics of the underlying, and therefore, if we knew the
parameter vector $\theta$ driving the underlying, we could price
options correctly. In this section, the input $\theta$ can be
$\xi$ as in the previous section or any more general parameter
governing the probability model for the underlying model.

A calibration method for an option pricing model must not rely on an
existing and liquid option's market, as otherwise option prices can
only be defined if they already exist. Therefore, we will assume
that there is not any existing option's market for that underlying
to use as a reference, and to determine, in this case, optimal
decisions for pricing.

We assume that as market makers, we must determine the best option
price to quote (which in principle needs not be its value), around
which we are to add a spread for bid/ask market making. In principle
we can assume that this spread is constant and symmetric around that
value, so it can be ignored for utility computation purposes, as it
becomes a constant for every trade. Therefore, assume that the
trader will bid/offer options at the same level.

Define $\theta$ as the set of inputs driving the underlying, whose
true value is unknown. Let $\theta^M$ the set of implied inputs at
which we end up making a market, and $P(\theta)$ as the price of
an option under any given parameter set $\theta$. The utility
function of a buyer of such an option can be defined as
$U_B(P(\theta^M),P(\theta))$, while the utility function for a
seller can be defined as $U_S(P(\theta^M),P(\theta))$. They could
be asymmetric (eg better to overprice than underprice if I would
rather buy than sell the option) given the asymmetry of the payoff
or other aspects, like the current portfolio or views of the
trader, or limited risks he is allowed to undertake. We are
defining utility functions at the time the decision is made. We
are not considering the true value of the option as a function of
the (still unknown) maturity price or path, but instead as a
function of the inputs driving the underlying since we estimate
them at time t.

Since the true value of $\theta$ is unknown to us, we have a
probability distribution, $\pi(\theta)$ representing our
knowledge/uncertainty about its true value. This measure of
uncertainty is assumed to be accurate. To proceed, we maximize our
expected utility given the information available, represented by
$\pi(\theta)$. As a market maker, we do not know a priori whether
we are going to be buying or selling the option. Suppose that with
probability $p$ we sell, and with probability $1-p$ we buy, then
our utility function, as a function of $\theta$ is equal to
$U(P(\theta^M),P(\theta)) = p*U_S(P(\theta^M),P(\theta)) +
(1-p)*U_B(P(\theta^M),P(\theta))$, for each possible value of the
true unknown parameter $\theta$. Our target is, therefore, to find
the optimal value for $\theta^M$ that maximizes
$E_{\pi}[U(P(\theta^M),P(\theta))]$ with respect to the
probability distribution $\pi(\theta)$. The optimal $\theta^M$
varies depending on our utility function, pricing model P and
posterior distribution $\pi$.
\subsection{Utility functions}
\begin{center}
\begin{figure}[t!]
\epsfig{file=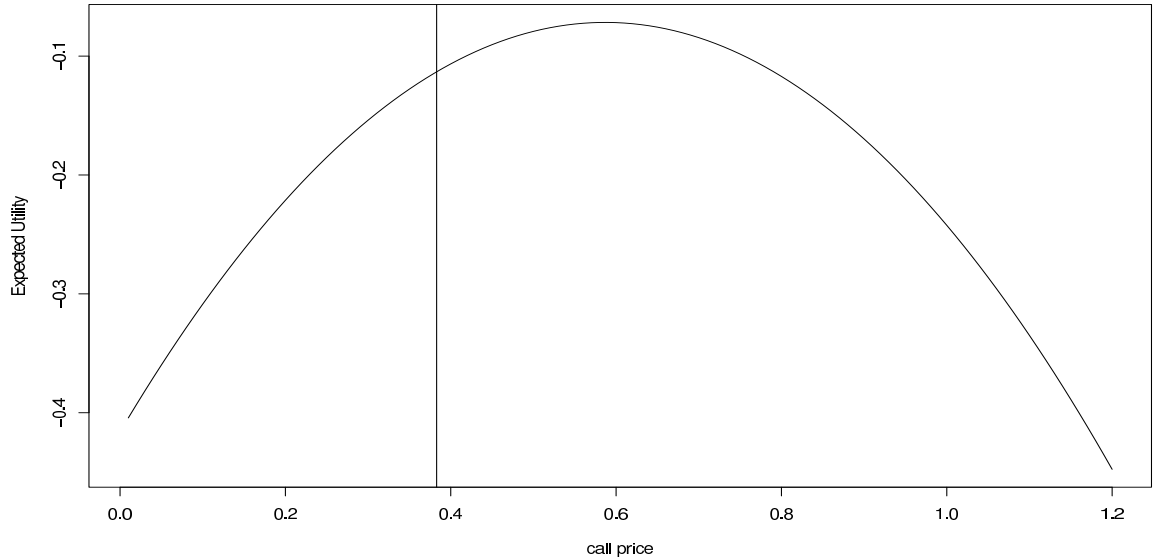,width=6in,height=3in}
\caption{Expected utility under each call price (curve) using the
posterior $\pi(\theta)$ versus call price and corresponding
utility if we were to use the most likely value of the parameter
(vertical line).} \label{fig:EXPECTED_UTILITY_VERSUS_CALL_PRICE}
\end{figure}
\end{center}

\begin{itemize}
\item {\bf 0-1 Utility function} When our utility is $1$ if
$P(\theta^M)=P($true $\theta)$, and 0 otherwise, we maximize it by
hitting the true value of the parameters (assuming a one-to-one
relationship between parameters and outputs, which is not
necessarily the case). In this case the optimal solution is
achieved when $\theta^M = \hat{\theta}$, which is our most likely
value. The optimal decision would be to value all options using
the most likely set of inputs. Unfortunately, given the
unobservability of $\theta$, the utility can not be quantifiable.
Plus, under a continuous $\theta$, we know that we reach the
maximized utility with probability zero, so the operational
exercise is futile.

\item {\bf Market volatility utility function} Say we want to
become a market-maker only if we have enough knowledge regarding
the market value of the option. Then our utility function could be
written as $U((P(\theta^M),P(\theta))$ if
$\sigma_{P(\theta)}\leq{\text{threshold}}$, and $0$ otherwise. The
optimal decision will depend not on a single input value, but
instead on the uncertainty about the potential output values, as,
if the uncertainty is too large, the optimal decision is not to
make a market (or make it at a different spread). The utility is
purely based on a measure of the input's propagated uncertainty,
instead of a single input estimate.

\item {\bf General utility functions} In general, the optimal
parameter set would be the one that maximizes some risk-averse
utility function, that is $\theta^M = argmax_\theta
E_{\pi}[U(P(\theta^M),P(\theta))]$.\\

We show an example in Figure
\ref{fig:EXPECTED_UTILITY_VERSUS_CALL_PRICE}. We assume a simple
quadratic utility function that penalizes divergence of the price
we quote, from the true (unknown) value implied by the true
(unknown) $\theta$, that is $U(P(\theta^M),P(\theta)) =
-[P(\theta^M)-P(\theta)]^2$. Furthermore, our knowledge of
$\theta$ will be represented through a gamma density
$\pi(\theta)\sim{Ga(\theta|2,1)}$. Using Black $\&$ Scholes for
pricing an at-the-money, 1-period maturity call, assuming the
risk-free rate is equal to zero, and strike price at 1, the
pricing function becomes a simple expression of the volatility
parameter $\theta$, namely
$P(\theta)=\Phi(\theta/2)-\Phi(-\theta/2)$, where $\Phi$ is the
normal cumulative distribution function. In this case, a maximum
of the expected utility is obtained at $P(\theta^M)$ around 0.59,
which corresponds to $\text{argmax}_{\theta^M}
\int{[-[P(\theta^M)-P(\theta)]^2] * \pi(\theta) d\theta}$.

This optimal price is neither the price at the expected value of
the parameter, P($E_{\pi}(\theta)$=2)=0.68 nor the price at its
most likely value, P($\hat{\theta}$=1)=0.38. The optimal and
expected prices are different for different distributions of
(estimates of uncertainty about) the inputs $\pi(\theta)$,
therefore, a full measure of the input uncertainty will affect the
optimal option value to use for decision-making purposes.
Parameter uncertainty assessment and propagation to the final
output would allow the market-maker to adjust market valuations
according to his own views on the probability of being bid/offer
$p$ and his own views/utility function about the parameter, in a
more systematic way. Notice that an easy extension would be to
consider $p$ random and perform the previous integral also over
$p$ as well, according to some probability distribution for this
parameter, which would incorporate the market maker's uncertainty
regarding the side that the counterparty takes. We must
incorporate full input uncertainty into output uncertainty if we
are to target optimal decision-making procedures.
\end{itemize}

We must stress that what we are computing probability distributions
and credible intervals for the risk-neutral option value given the
information available at each point in time, and not credible
intervals for any actual discounted payoff of that option, which
remain stochastic.

\section{Likelihood and model equations}\label{foundations}
Cox et al., (1979) proposed a tree model for valuing options. We
propose a new method to calibrate their model to asses the
propagation of uncertainty.
\subsection{The Probability Model}
We start with the classical binomial model, where the value of the
underlying at time $t$, $S_{t}$, follows dynamics as defined in
Cox et al., (1979):
\begin{eqnarray}
S_{t+1} &=& S_{t}\xi
\end{eqnarray}
where $\xi$ is a dichotomic random variable that represents the
possible up and downs of the underlying (only those two moves are
assumed possible), and that generates a whole binomial tree, with
the following probability distribution:
\[ \xi = \left\{\begin{array}{ll}
u & \mbox{with probability $p$}\\
d & \mbox{with probability $1-p$.}
\end{array}
\right. \] where $u$ is the movement of the underlying in the up
direction, and $d$ is the movement of the underlying in the down
direction.\\

In practice, $\xi$ is unobservable, and so are the values $u$ and
$d$. However, conditional on $u$ and $d$, the value of $\xi$ is
specified and a whole binomial tree is specified as well, allowing
us to price options. Calibration methods using observed option
prices are typically used to obtain the values
of $u$ and $d$ implied by the market under the model.\\

In most cases, however, options market prices could be unreliable,
the options market could be underdeveloped or simply options price
discovery could prove expensive or unfeasible. Additionally, they
include market risk premia, which distorts calculations if the
target is the theoretical, risk-neutral value. Furthermore an
options market needs not exist, especially for ad-hoc or
client-specific options. In these cases, calibration is not
possible, and the most we should expect is to extract some
information about potential $u$ and $d$ from the historical return
series of the underlying, to value an option on it.\\
Let $R_i$ for $i=1,\dots{n}$ be the returns of the underlying over
the $n$ periods in our sample, where $n$ could be, indeed, quite
small. In this case we can only attempt to use this model to price
options by extracting $u$ and $d$ from this sample. Therefore what
we observe are noisy realizations of $\xi$, which we denote as
$\xi_i = 1+R_i = r_{i}$. Our aim is to extract from the observed
$\xi_i$ information about the underlying process to properly value
options on that underlying, as well as to account for the
uncertainty about that process. Then one can update the tree in a
sequential manner as soon as more realizations of $\xi$ are
observed.

The first problem we face is that we cannot assume to know (or
estimate without uncertainty) the value of the $u$ and $d$ with
which to generate the binomial tree. Accounting for the level of
parameter uncertainty is vital for proper practical option pricing.
This is even more important in the case of undeveloped options
markets, where spreads tend to be sometimes unreasonably large due
to the uncertainty felt by market makers as not being accounted for
in the theoretical (point estimate) values.

Additionally, option uncertainty will reflect a skewed nature, as we
will show, so the most likely price could be quite different from
the expected price or even from the optimal price under a certain
loss function (for example drawdown-based loss functions). All these
elements lead us to consider computing not only a calibrated
parameters, but the overall uncertainty we have about them.

Assuming that the CRR model is true, we would, therefore, observe
$n$ values of $\xi_i$ that could potentially have been the true
value of some underlying $\xi$ generating the process.
Additionally, we know that the no-arbitrage condition must be met,
and, therefore, $\xi_i$ under the down moves must be between 0 and
$1+r_f$, and $\xi_i$ under the up moves must be between $1+r_f$
and $\infty$. One might have the temptation to model the $\xi_i$
as a simple mixture of normals or any other overlapping mixture.
This would intrinsically violate the no-arbitrage condition, as
nothing would prevent the $\xi_i$ under the up moves to be smaller
than the $\xi_i$ under the down moves. We need a statistical
approach that accommodates to the restrictions (truncation) in the
pricing model for the up and down returns.

In summary, if we want to extract the information contained in the
series $\xi_i$ about the generating process of an underlying, and
additionally we assume the CRR pricing mechanisms are proper for
pricing this underlying, a natural choice would be a mixture of
non-overlapping (truncated) densities for the up and down moves.
We consider the simplest of these mixtures in our formulation,
that is the mixture of Truncated Normal densities. For simplicity,
we also assume that the generating process is constant in time,
although this assumption could easily be relaxed.

Our choice for a parametric formulation, as opposed to, for
example, Monte-Carlo/bootstrap-based methods, is that the observed
range of data could easily be much more narrow than the potential
range, leading to underestimation of the tails/extreme values that
could eventually happen. This is of key importance when the
pricing tools are applied to risk management, as measures like VaR
and expected shortfall would be heavily affected by a correct
accounting for potential extreme values.

Our goal, therefore, is to update our knowledge of some unknown
potential value of $\xi$ that describes the future moves of the
underlying (and the corresponding option prices), but accounting
for the inherent
uncertainty and variability of $\xi$.\\

\subsection{\protect\bigskip Risk neutral measure for the CRR model}
Given a $\xi$, that is, given a value of both $u$ and $d$, we can
generate a whole binomial tree, and in order to rule out arbitrage
and under the condition of market equilibrium in any node of the
tree, the expected value of the underlying at the end of a given
node $t$ in the binomial tree is $E_{q}\left[
S_{t+1}|\mathcal{F}_{t}\right] =S_{t}(1+r_{f})$, where $S_{t}$ is
the underlying price at the beginning of the node $t$, and $E_{q}$
denotes expected value with respect to the risk neutral measure
$q$. In our model, this last condition is formalized as:
\begin{eqnarray}
S_{t}(1+r_{f}) &=& qS_{t}u+(1-q)S_{t}d
\end{eqnarray}

solving for $q$ yields:

\begin{equation}\label{rnq}
q=\frac{(1+r_{f})-d}{u-d}
\end{equation}

\subsection{Description of other classical methods}
In this section we present alternative methods to ours. In the
next section we present our method. Both of the
alternative methods we describe aim at reconstructing the
probabilistic nature of a recombinant tree describing the dynamics
of the underlying from the observed option prices.
\begin{itemize}
\item {\bf Rubinstein's method} It is based on the following
natural assumptions:
\begin{enumerate}
\item The underlying asset follows a binomial process.
\item The binomial tree is recombinant.
\item Ending nodal values are ordered from lowest to highest.
\item The interest rate is constant.
\item All paths leading to the same ending node have the same
probability.
\end{enumerate}
Once the probabilities and returns of the final nodes are
established, the recombinant nature of the binomial tree plus the
non-arbitrage condition, are used to inductively obtain the
probabilities of arrival at the previous set of nodes along the tree
as well as the returns at these nodes.

The implied posterior node probabilities are obtained solving the
following quadratic program:
$$\min\{\sum_{i=1}^n(q_i - q_i')^2 \,|\,(q_1,...q_n) \in \mathcal{C}\},$$
\noindent where $\mathcal{C}$ is the following set of constraints:
$$
\begin{array}{rcl}
&&(i)\;\; q_i \geq 0,\;\;\sum q_j = 1,\\
&&(ii)\;S^b \leq \sum q_jS_j/(1+r_{f})^n \leq S^a,\\
&&(iii)\,P(\theta)_i^b \leq \sum q_j(S_j - K_i)^+ \leq
P(\theta)_i^a, \;\;\;\;i = 1,...,m.
\end{array}
$$
Here $S^a, S^b$ are respectively the current ask and bid prices of
the underlying asset. $P(\theta)_i^a$ and $P(\theta)_i^b$ are the
prices of the European calls on the assets with strike prices
$K_i$ for $i = 1,...,m.$ The $S_j$ are the end nodal prices and
the $q_j'$ are some prior set of risk neutral probabilities
associated with the given tree.

\item {\bf Derman and Kani's procedure}

Their aim is to understand how the underlying must evolve in
such a way that the prices of the European calls and puts coincide
with the market prices for the different strike prices.

The method ends up with the construction of a recombinant tree for
which the up and down shifts and the transition probabilities may
change from node to node at each time lapse.

If the prices at the nodes at time $t_n$ have been reconstructed,
they propose the following routine for reconstructing the prices at
time $t_{n+1}$ as well as the transition probabilities out of each
node at time $t_n$.

If $s_i$ is the stock price at the $i$-th node at time $t_n$, then
the end-prices at time $t_{n+1}$  are $S_{i+1} > S_i$ and the
probabilities $P_i$ of such a move are reconstructed from a risk
neutral requirement and a matching of the option price at time
$t_{n+1}$ as if strike price were $s_i$.

This procedure provides $2n$ equations of the $2n + 1$ needed to
for for determining the $n + 1$ prices and the $n$ transition
probability. They propose a centering condition, to make the
center coincide with the center of the standard CRR tree, to close
the system of equations.
\end{itemize}

In both methods outlined above, one ends up with a recombinant
tree, describing the dynamics of an underlying in which the prices
and/or transition probabilities may change from time to time and
from node to node. Once the tree is at hand, one may carry out the
computation of all quantities of interest.

In our approach, the historical record of up and down prices of the
underlying is used to update a standard binomial tree describing the
time evolution of the asset, except that what we have is a whole
collection of trees, each occurring with a posterior probability. For
each of these trees, we can compute whatever we want, for example
the price of a European call, except that each of these values has a
posterior probability of occurrence.

\subsection{The Statistical Modelling and Estimation}\label{stats} For the purpose of our estimation,
we assume that we observe independent, equally-spaced realizations
of $\xi_i$ from the past. Then the joint likelihood is just but the
product of $n$ independent realizations:
\begin{equation*}
L(\xi_{1},...,\xi_{n}|u^*,\sigma _{u}^{2},d^*,\sigma _{d}^{2},p)=\overset{n}{%
\underset{i=1}{\Pi }}\left(
pTN(\xi_i|u^*,\sigma_{u}^{2},1+r_{f},+\infty)+(1-p)TN(\xi_i|d^*,\sigma_{d}^{2},0,1+r_{f})\right)
\end{equation*}
We assume that the distribution of $\xi_i$ under the up moves is a
truncated normal with parameters $u^*$ and $\sigma_u$, which
happens with unknown probability $p$, while the distribution of
$\xi_i$ under the down moves is also a truncated normal with
parameters $d^*$ and $\sigma_d$, which would happen with unknown
probability $1-p$. This is consistent with the formulation in
(4.1), but we basically acknowledge our uncertainty and natural
variability about the potential values under the up/down moves.
Later we show how to transfer this into uncertainty about the options prices.\\

In a Bayesian framework, together with the likelihood, we complete
the joint distribution with the following priors for our
parameters:
\begin{eqnarray}
d^* &\sim &U(d^*|0,1+r_{f}) \\
u^* &\sim &U(u^*|1+r_{f},2) \\
\pi (\sigma _{u}^{2}) &\sim &IG(\sigma _{u}^{2}|\alpha _{u},\beta _{u}) \\
\pi (\sigma _{d}^{2}) &\sim &IG(\sigma _{d}^{2}|\alpha _{d},\beta _{d}) \\
\pi (p) &\sim &Beta(p|a,b)
\end{eqnarray}

Also notice that we define the location parameters as $u^*$ and $d^*$, since these location parameters are not the expected
values of the distributions under the up/down moves respectively.\\
Finally, to match our notation to that of CRR, we define $u$ as a random variable with density that of $\xi$ under the up moves,
and $d$ a random variable with density that of $\xi$ under the down moves. In summary,

\[ \xi = \left\{\begin{array}{ll}
u \sim{} TN(u|u^*,\sigma_{u}^{2},1+r_{f},+\infty) & \mbox{with probability $p$}\\
d \sim{} TN(d|d^*,\sigma_{d}^{2},0,1+r_f) & \mbox{with probability
$1-p$.}
\end{array}
\right. \]

We chose a mixture of truncated gaussian for several reasons. First we wanted to mimic the assumptions of the options
model we use, which force the existence of two kinds of
observations: positive, defined as those above the risk-free, and
negative, defined as those below the risk-free. This naturally
brings the idea of a mixture of truncated densities. However, the
inherent uncertainty about the values of those, together with the
fact that in reality we don't simply observe two single values for
the series, suggests that the realized positive and negative
values do come from some mixture, with the truncation at the point
of division between the ups and downs (the risk-free). We allowed
different variances for the positive and negative parts to account
for possible skewness in the data, potential bubbles and other
non-symmetric market behaviors. Finally our choice of gaussian
distributions just came from trying to keep the choices simple.

We could in principle have added further layers of complication to
the model (e.g. mixtures of t-distributions, markov switching
behavior or even stochastic volatility or jumps), but decided to
keep things simple, as our target, again, is not a better pricing
model, but a method for a more accurate description of the involved
uncertainties. All these extensions are tractable, because the
statistical model is fully constructed on the physical measure, and
plenty of estimation algorithms are available for these potential
extensions. The extent of divergence in the final output is,
indeed, an interesting topic in itself, but we will focus here on procedures for propagating uncertainty.\\

It is worth noting that this truncated normal approach becomes a single gaussian distribution in
the limit (as $u^\ast$ gets close to $d^\ast$, and both to $1+r_f$, $p$ gets close to 0.5 and $\sigma_u$ gets close to $\sigma_d$).
Also we can see that as $\sigma_u$ and $\sigma_d$ get close to zero, the model is equivalent to the original
CRR framework, as we have two point masses at $u^\ast$ and $d^\ast$.

We use a Markov Chain Monte Carlo (MCMC) approach to estimate the
parameters in the model. Details of the actual algorithm are
outlined in the Appendix.

\subsection{Monte Carlo use of the MCMC output}\label{methods}
In the following subsections, we present three alternative methods
for calibration that we shall refer to as the $\theta$ method, the $\xi$ method and
the expected $\xi$ method. This names will become clear in the next subsections.
\subsubsection{The $\theta$ method}
The approach described in the previous section allows us to assess
the level of uncertainty in the inputs of the option pricing
formula. Our posterior distribution for the parameters
$\theta=[u^\ast,d^\ast,\sigma_u^2,\sigma_d^2]$ given the data
represents our uncertainty about the parameters driving the dynamics of the underlying, where each combination of
values has associated a certain likelihood under the posterior
distribution.\\
The next step is to use those values to generate possible option
prices. A Monte Carlo approach suffices here to propagate the
uncertainty of the inputs into uncertainty about the outputs
(option prices). The idea is to draw random samples from the
posterior distribution, generate potential up and down moves, and
pass them to the option pricing formula to obtain (random)
possible outputs. This will effectively provide us with the
posterior distribution of the option prices given the
data.\\
In our problem, what we are trying to compute is the following
double integral, where data$=\left\{
u^{1},...,u^{l},d^{1},...,d^{k}\right\} $, where $l+k=n$, and
since $P(\xi )$ only depends on $(u,d)$, it is independent of
$\theta $ given $\xi =(u,d)$:

\begin{eqnarray*}
P|\text{data} &=&\int_{\Theta }\left[ \int_{\Xi }P(\xi )f(\xi |\theta )d\xi %
\right] \pi (\theta |Data)d\theta  \\
P|\text{data} &=&\int_{\Theta}E_{\xi |\theta }\left[ P(\xi
)\right] \pi (\theta |Data)d\theta
\end{eqnarray*}

where $P|\text{data}$ is the risk neutral price conditional on the data available and averaged
over all the values of $\theta$.
The quantity $E_{\xi |\theta }\left[ P(\xi )\right] $ is a function of $%
\theta $, averaged over the different $\xi|\theta$. We refer to $E_{\xi |\theta }\left[ P(\xi )\right] $ as
the option price given $\theta$. Therefore, we need to integrate
out $\theta $ from every $E_{\xi |\theta }\left[ P(\xi )\right] $
by integrating $E_{\xi |\theta }\left[ P(\xi )\right] $ with
respect to the posterior of $\theta $. The pseudo-code for the
Monte Carlo part of this first method is comprised by the
following steps:\\
\begin{enumerate}
\item Draw (uniformly) a sample from the MCMC output $\theta_i=\{u^\ast,d^\ast,\sigma_u^2,\sigma_d^2\}$.
\item From $\theta_i$, generate a sequence $\left\{(u_{i},d_{i})\right\}_{i=1}^{M}$
of possible values for $u$ and $d$ using $f(\xi|\theta)$.
\item Use these values of $u_{i}$ and $d_{i}$ to
generate $q_{i}$ from equation \eqref{rnq} and the corresponding
$\xi_{i}$:
\[ \xi_{i} = \left\{\begin{array}{ll}
u_{i} & \mbox{with probability $q_{i}$}\\
d_{i} & \mbox{with probability $1-q_{i}$}
\end{array}
\right. \] Compute $\left\{P(\xi_{i})\right\}_{i=1}^{M}$ which is a
sequence of binomial trees for that given $\theta$.
\item Go back to step 1 and
repeat $L$ times in order to generate a set of averages.
\item If we want a single price $P|\text{data}$, then average these averages.
\end{enumerate}

\subsubsection{The $\xi$ method}
Here we draw random samples from the posterior distribution
of $u^\ast,d^\ast,\sigma_u^2,\sigma_d^2$, generate potential up
and down moves through $\xi$, and pass them to the option pricing
formula to obtain (random) possible outputs. This will effectively
provide us with the posterior distribution of the option prices
given the data. There are two possible outcomes that we are
interested in. First we are interested in the overall distribution
of the option price given the data, as we motivated in subsection
\ref{uncertainty} and section \ref{utilities} In this case, we
need to draw a sequence of $\theta_{i}$ form the posterior
distribution $\pi(\theta|\text{data})$ in order to generate as
many Monte Carlo samples from $f(\xi|\theta)$. Finally, since we
do not know analytically the predictive marginal posterior
distribution $f(\xi|\text{data})$, we need to partition the values
of $u$ and $d$ into many bins and compute the proportion of
$\xi$'s falling in each bin.
The option price is computed by solving the following double
integral:
\begin{eqnarray*}
P|\text{data} &=&
\int_{\Theta}\left[\int_{\Xi}P(\xi)f(\xi|\theta)d\xi\right]\pi(\theta|\text{data})d\theta\\
P|\text{data} &=&
\int_{\Xi}P(\xi)\left[\int_{\Theta}f(\xi|\theta)\pi(\theta|\text{data})d\theta\right]d\xi\\
P|\text{data} &=& \int_{\Xi}P(\xi)\pi(\xi|\text{data})d\xi
\end{eqnarray*}
where $\theta\equiv [u^\ast,d^\ast,\sigma_u^2,\sigma_d^2]$,
$P(\xi)$ is the option price generated by a tree given
$\xi=(d,u)$. Given that we do not know analytically
$\pi(\xi|\text{data})$, we can approximate it by generating a
sequence $\left\{\xi_{i}\right\}_{i=1}^{L}$ for $L$ big enough,
and construct $M$ bins with center
$\bar{\xi_{k}}=(\bar{u_{k}},\bar{d_{k}})$ in order to approximate
$f(\xi|\text{data})\approx\sum_{k=1}^{M}p_{\bar{\xi_{k}}}\delta^{\bar{\xi_{k}}}$,
where $p_{\bar{\xi_{k}}}$ is computed as the number of
elements\footnote{$\delta^{\bar{\xi_{k}}}$ is the Dirac delta
function.} in $\left\{\xi_{i}\right\}_{i=1}^{L}$ that fall in the
$k$-th bin with center $\bar{\xi_{k}}$.

We now propose a pseudo-code for the Monte Carlo that is comprised by the following steps:\\

\begin{center}
\begin{figure}[t!]
\epsfig{file=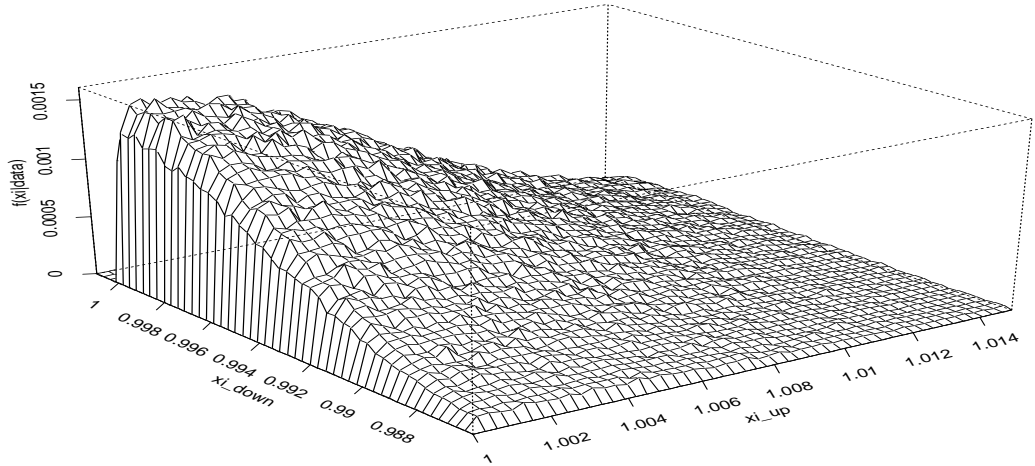,width=6in,height=5in}
\caption{Posterior probability distribution for $\xi$ given the
data using the $\xi$ method.} \label{fig:POSTERIOR_MODEL}
\end{figure}
\end{center}

\begin{enumerate}
\item Draw (uniformly) a sample from the MCMC output
$\theta_i=[u^\ast,d^\ast,\sigma_u^2,\sigma_d^2]$. \item Generate $M$
possible values for $\xi=(u,d)$, given
$\theta=[u^\ast,d^\ast,\sigma_u^2,\sigma_d^2]$, using
$f(\xi|\theta_i)$.
\item Iterate many times steps 1 and 2 in order
to generate a sequence $\left\{\xi_{i}\right\}_{i=1}^{L}$. \item
Use $\left\{\xi_{i}\right\}_{i=1}^{L}$ generated in steps 1
through 3 to approximate
$\pi(\xi|\text{data})\approx\sum_{k=1}^{M}p_{\bar{\xi}}\delta^{\bar{\xi}}$
with the use of bins. \item Sample $\xi_{i}=(u_{i},d_{i})$ with
probability given by $\pi(\xi_{i}|\text{data})$ for $i=1,...,L$
computed in the previous step, and construct its associated tree
given by:
\[ \xi_{i} = \left\{\begin{array}{ll}
u_{i} & \mbox{with probability $q_{i}$}\\
d_{i} & \mbox{with probability $1-q_{i}$}
\end{array}
\right. \] $\xi_{i}$ generates a binomial trees that allows us to
compute an option value unconditional of $\theta$.
\end{enumerate}
It is worthwhile noticing that the posterior distribution
$\pi(\xi_{i}|\text{data})$ quantifies model uncertainty regarding
the space of all available tree models for the underlying. Indeed,
each $\xi_{i}$ parameterizes a whole tree model which has its
associated probability given by $\pi(\xi_{i}|\text{data})$ (Cont,
R. (2006)). The posterior probability distribution of each model
is given in figure \ref{fig:POSTERIOR_MODEL}. As the amount of
data increases, the posterior distribution $\pi(\xi|\text{data})$
becomes tighter as can be seen in figure \ref{fig:POSTERIOR_MODEL
DATA}.

\begin{center}
\begin{figure}[t!]
\epsfig{file=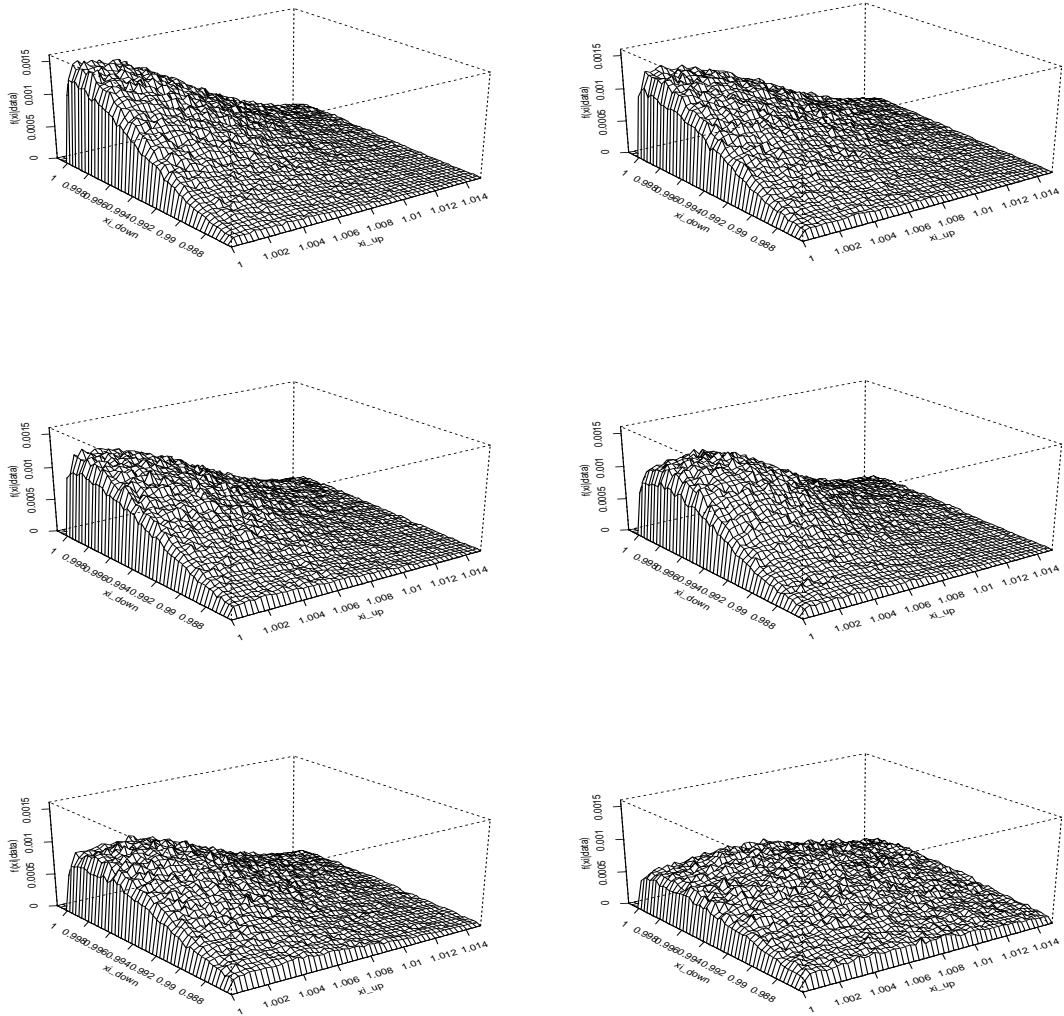,width=6in,height=6in}
\caption{Posterior probability distribution for $\xi$ given the
data using the $\xi$ method. Plots are for 252, 126, 65, 21, 15 and 10 business days from
left to right, top to bottom, respectively.}
\label{fig:POSTERIOR_MODEL DATA}
\end{figure}
\end{center}

\subsubsection{The expected $\xi$ method}
In this subsection we present the third and last method that
consists of using the expected value of $\xi$ as inputs for the
tree model. The idea behind this derivation can be found in Cox et
al., (1979), where $u$ and $d$ are known with probability one.
However, as these are model parameters, they can only be partially
observed together with some noise. Therefore, it is the expected
value of $u$ and $d$ and not their realization which is the
quantity of interest under this last approach. What is therefore
observable is $u=E(u)+\epsilon_{1}$ and $d=E(d)+\epsilon_{2}$,
where $\epsilon_{1}$, and $\epsilon_{2}$ are normally distributed
random variables. The idea of this method is to propagate the
uncertainty of the parameters
$[u^\ast,d^\ast,\sigma_u^2,\sigma_d^2]$ through both $E(u)$ and
$E(d)$ as inputs for the tree model. This method can be seen
similar to the Bootstrapped Mean method (BM) that will be
described in section \ref{classicalcalibration} The expected $\xi$
method method accounts for parameter uncertainty, whereas the
Bootstrap mean method does not, even so for small sample sizes.

The option price is computed by solving the following integral:
\begin{eqnarray*}
P|\text{data} &=& \int_{\Theta}P\left[E_{\xi |\theta
}(\xi)\right]\pi(\theta|\text{data})d\theta\\
P|\text{data} &=& \int_{\Theta}P\left[\int_{\Xi}\xi
f(\xi|\theta)d\xi\right]\pi(\theta|\text{data})d\theta
\end{eqnarray*}

We now describe the following pseudo-code for the Monte
Carlo:\\

\begin{enumerate} \item
Draw (uniformly) a sample from the MCMC output. The sample is a
vector of parameters $\theta=[u^\ast,d^\ast,\sigma_u^2,\sigma_d^2]$.
\item Given $\theta=[u^\ast,d^\ast,\sigma_u^2,\sigma_d^2]$, compute the
following moments\footnote{Given that $u\sim TN(u|u^{\ast },\sigma
_{u}^{2},1+r_{f},+\infty)$ and $d\sim TN(d|d^{\ast},\sigma
_{d}^{2},0,1+r_{f})$}:
\begin{eqnarray}
E(u) &=&u^{\ast }+\sigma _{u}\frac{\phi (c_{0})}{1-\Phi (c_{0})} \\
E(d) &=&d_{\ast }-\sigma _{d}\frac{\phi (b_{0})-\phi (a_{0})}{\Phi
(b_{0})-\Phi (a_{0})}
\end{eqnarray}

\item Compute the expected value of $\xi$
\[ E(\xi) = \left\{\begin{array}{ll}
E(u) & \mbox{with probability $q$}\\
E(d) & \mbox{with probability $1-q$}.\end{array}\right. \] \item
Compute $P[E(\xi)]$ \item Go to step 1 and repeat many times.
\end{enumerate}
Here $c_{0}=\frac{1+r_{f}-u^{\ast }}{\sigma _{u}}$, $%
a_{0}=\frac{0-d^{\ast }}{\sigma _{d}}$,
$b_{0}=\frac{1+r_{f}-d^{\ast }}{\sigma _{d}}$, $\phi $ is the
standard normal density and $\Phi $ is the standard normal cdf.
The risk neutral probability $q$ is equal to:
\begin{equation}
q=\frac{1+r_{f}-\left[ d^{\ast }-\sigma _{d}\frac{\phi
(b_{0})-\phi (a_{0})}{\Phi (b_{0})-\Phi (a_{0})}\right] }{u^{\ast
}+\sigma _{u}\frac{\phi (c_{0})}{1-\Phi (c_{0})}-\left( d^{\ast
}-\sigma
_{d}\frac{\phi (b_{0})-\phi (a_{0})}{\Phi (b_{0})-\Phi (a_{0})}%
\right) }
\end{equation}

\section{Empirical results}\label{results}

\subsection{S\&P500 Application}
We apply the three methodologies from the previous section to
determine the uncertainty in the price of call
options\footnote{USD LIBOR is used as the risk-free interest rate
as suggested by Bliss and Panigirtzoglou (2004).} for the S\&P500
for a period of 1993. We use this simple example, for which
pricing has become simple and almost automatic, to show the actual
uncertainty we have about the call prices due to uncertainty about
the pricing tool inputs. Again, we must stress that our target in
this paper is not to propose a better pricing tool, but instead to
show the effects of the uncertainty in inputs into option pricing,
and to show an example of how to propagate that uncertainty for a
simple
and well-known tree-based model.\\
Our data consists of daily returns for the S\&P500 index, from the
years 1992 and 1993. We will use, at time $t$, the returns of the
previous 252 business days, to estimate the parameters
$u^\ast,d^\ast,\sigma_u^2,\sigma_d^2$, using the procedures
outlined
in section \ref{methods} We show summaries of the convergence of the Markov Chains in the Appendix.\\
We use the posterior distribution of the parameters, together with
the underlying price at that time $S_t$ and for a strike of
constant $K=450$ to price an European call
option with maturity on Friday, December 17th, 1993.\\

\begin{center}
\begin{figure}[t!]
\epsfig{file=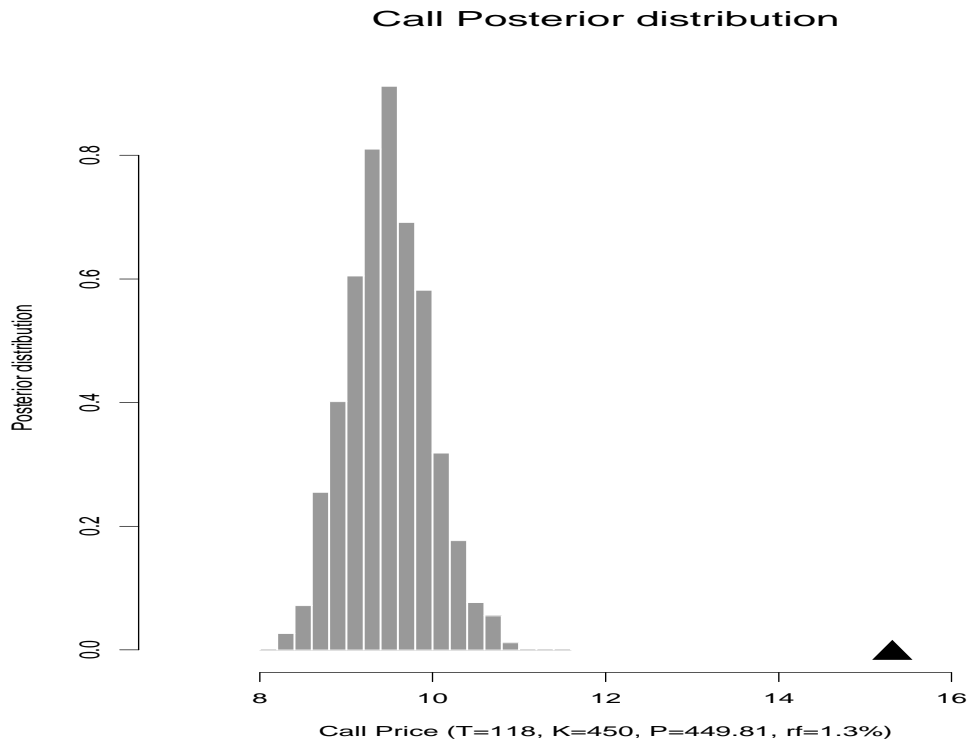,width=6in,height=4in}
\caption{Posterior density (histogram) under the $\theta$ method
and market price (triangle) for a call price for July 1st, 1993.}
\label{fig:POSTERIOR_CALL_PRICE_1}
\end{figure}
\end{center}

\begin{center}
\begin{figure}[t!]
\epsfig{file=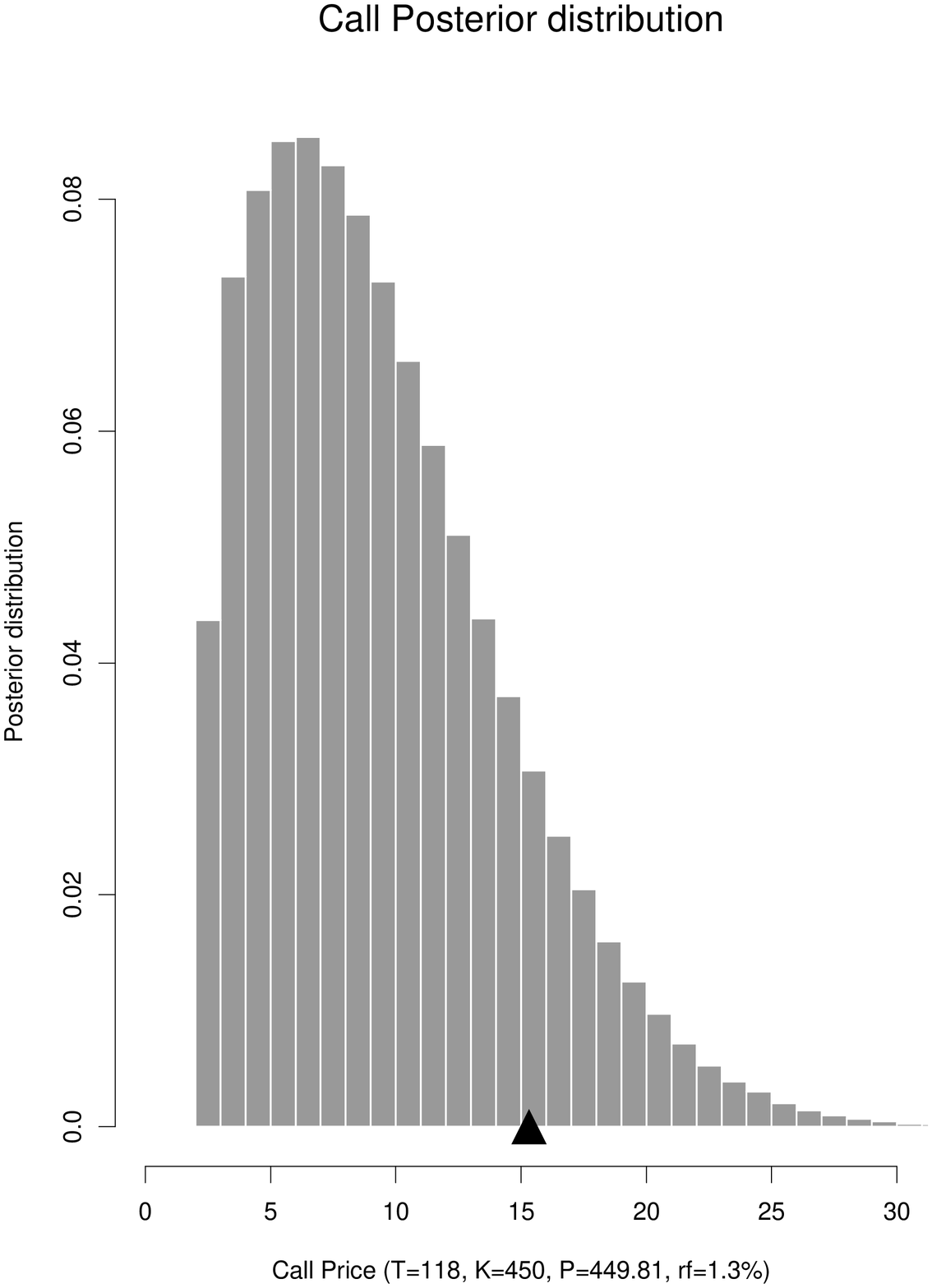,width=6in,height=4in}
\caption{Posterior density (histogram) under the $\xi$ method and
market price (triangle) for a call price for July 1st, 1993.}
\label{fig:POSTERIOR_CALL_PRICE_2}
\end{figure}
\end{center}

\begin{center}
\begin{figure}[t!]
\epsfig{file=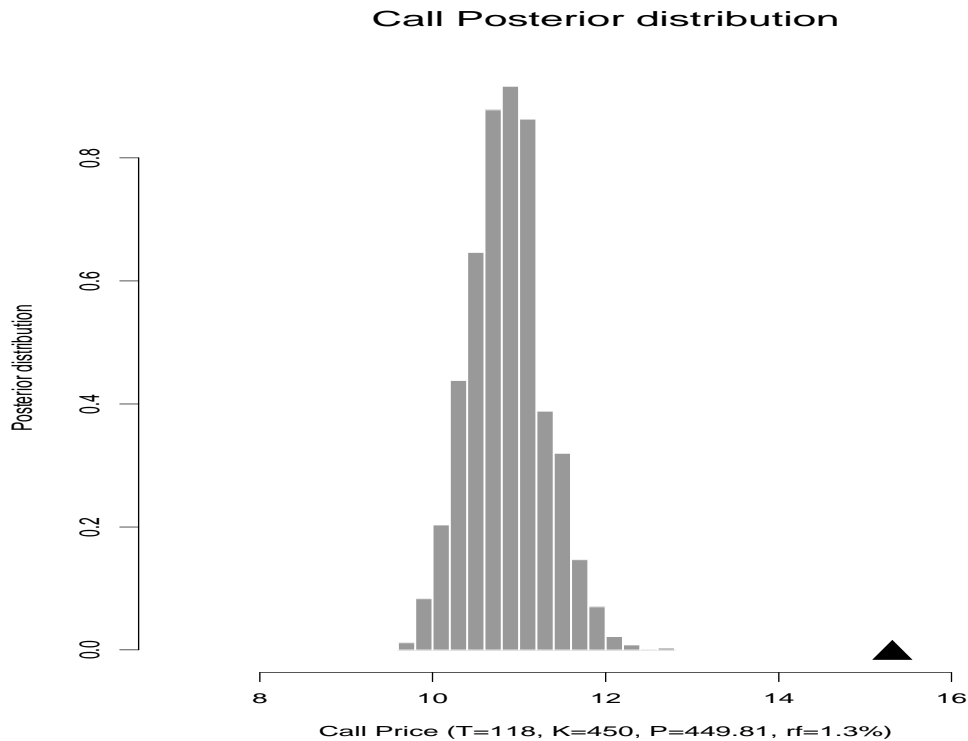,width=6in,height=4in}
\caption{Posterior density (histogram) under the expected $\xi$
method and market price (triangle) for a call price for July 1st,
1993.} \label{fig:POSTERIOR_CALL_PRICE_3}
\end{figure}
\end{center}

Figures \ref{fig:POSTERIOR_CALL_PRICE_1},
\ref{fig:POSTERIOR_CALL_PRICE_2}, and
\ref{fig:POSTERIOR_CALL_PRICE_3} show the posterior distribution
for the theoretical call value on July 1st, 1993 under the three
methods. The triangle represents the actual market price of the
option. There are three features that are worth noting. First, we
can clearly see that the posterior distribution from figure
\ref{fig:POSTERIOR_CALL_PRICE_2} is far from concentrated, which
is not the case for figures \ref{fig:POSTERIOR_CALL_PRICE_1} and
\ref{fig:POSTERIOR_CALL_PRICE_3}. The three figures reflect the
uncertainty about the inputs and how this propagates into
uncertainty about the call price output. Second, they show that
the call price is skewed, as also observed by Karolyi (1993). This
shows us that even the most likely value of the actual output is
not necessarily the most representative one. Under the three
methods we obtain similar results with different levels of
uncertainty as shown in table 1. Third, we should note that this
valuation is still being done under a risk-neutral approach, so
the actual call market price is (much) larger than one would
expect under risk neutrality. This by no means invalidates the
risk-neutral approach, but allows us a better perception and even
quantification of the extent of the risk premia in the market as
well as empirically the considerable overpricing (underpricing) of
in-the-money (out-of-the-money) options (MacBeth and Merville
(1979), Rubinstein (1985)). Our work differs from Karolyi (1993)
since our bayesian analysis is performed for a greater class of
stochastic processes as limits in continuous time, than the
classical geometric brownian motion as treated in Karolyi (1993),
which is a special case.
\begin{center}
\begin{figure}[t!]
\epsfig{file=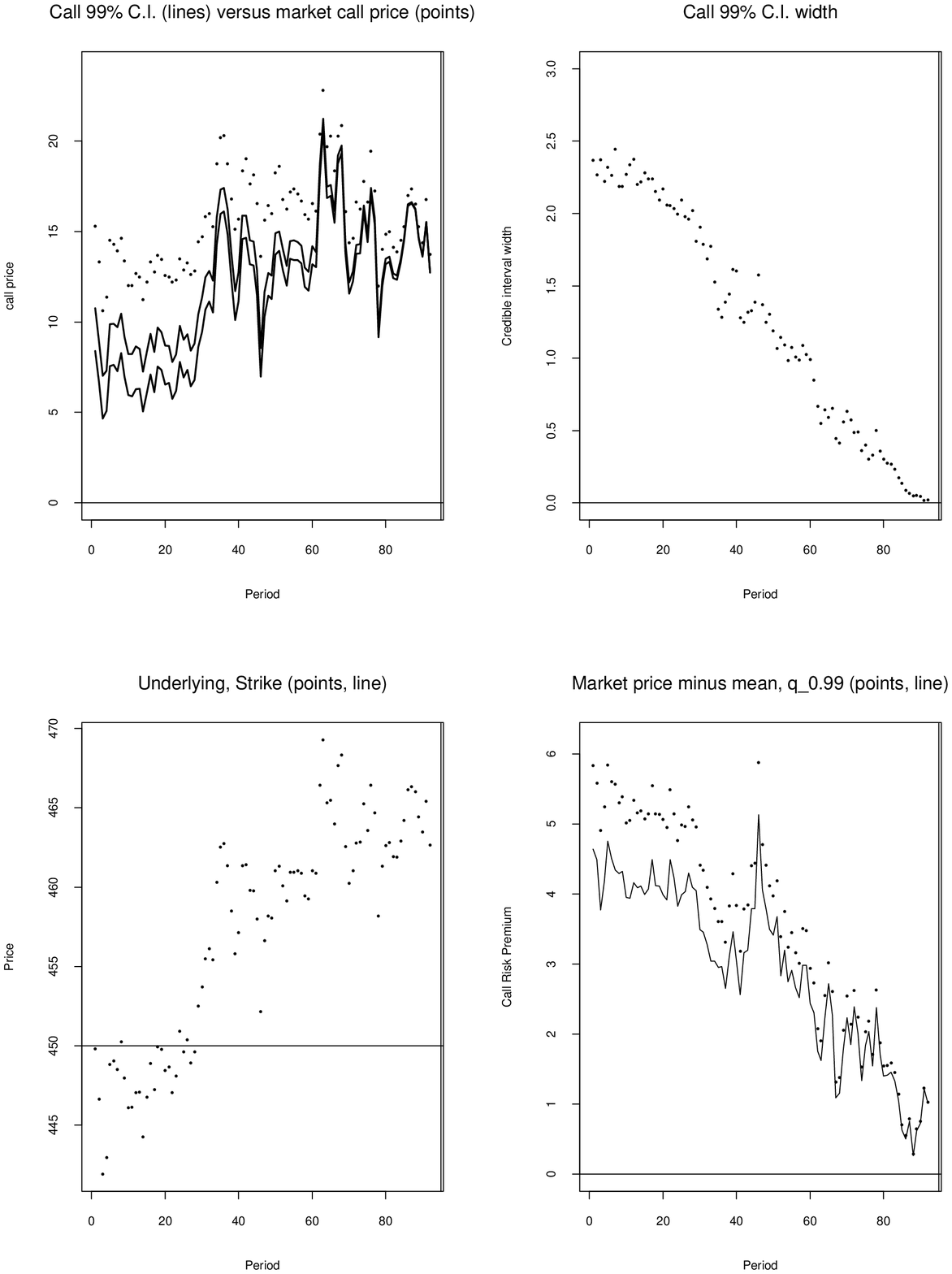,width=6in,height=4in}
\caption{99\% credible interval and market call prices (top-left),
99\% credible interval range (top-right), underlying level and
strike (bottom-left) and estimates of risk premium (bottom-right)
under the expected $\theta$ method} \label{fig:ALL_PLOTS_SP500_1}
\end{figure}
\end{center}

\begin{center}
\begin{figure}[t!]
\epsfig{file=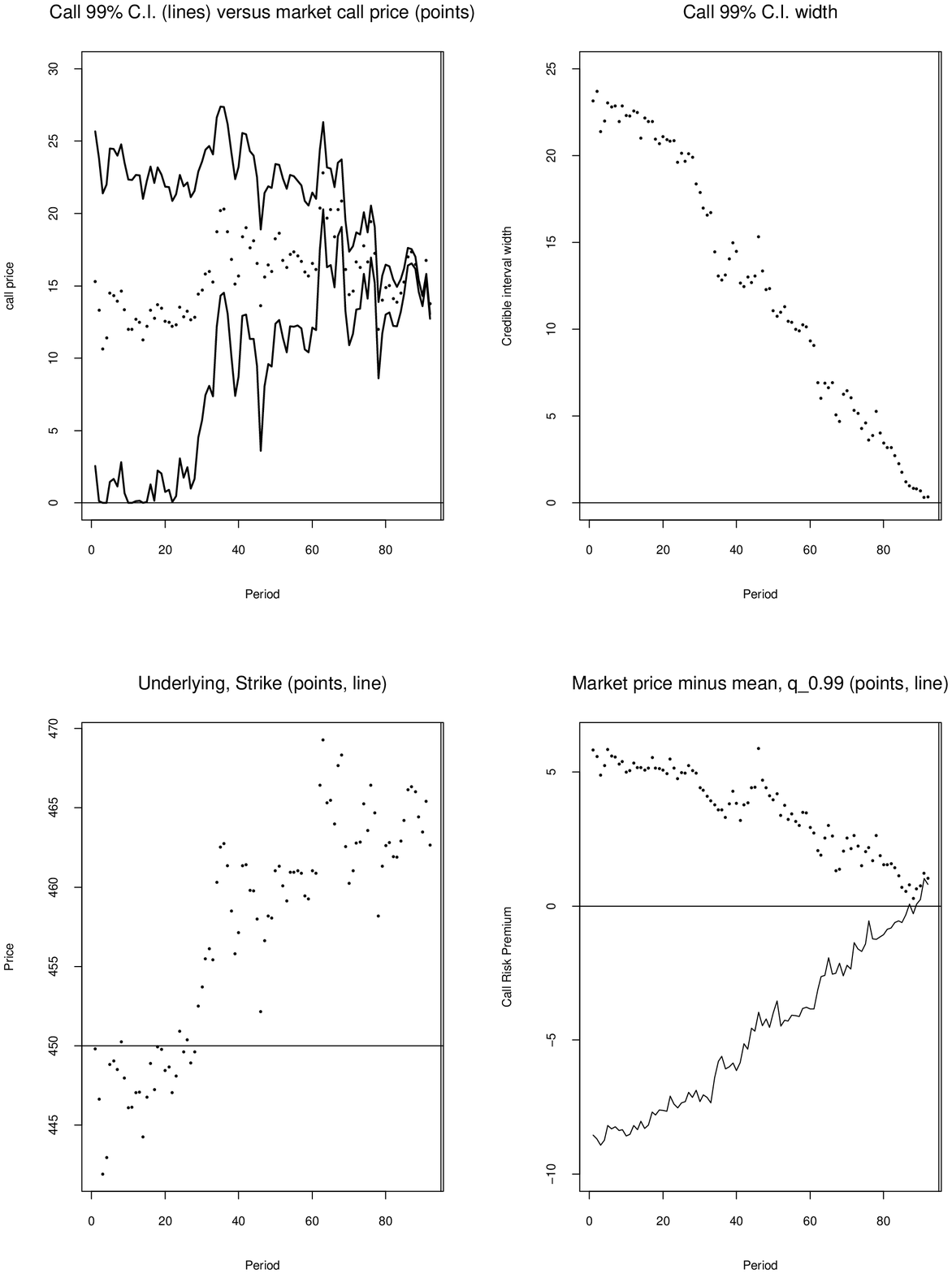,width=6in,height=4in}
\caption{99\% credible interval and market call prices (top-left),
99\% credible interval range (top-right), underlying level and
strike (bottom-left) and estimates of risk premium (bottom-right)
under the $\xi$ method} \label{fig:ALL_PLOTS_SP500_2}
\end{figure}
\end{center}

\begin{center}
\begin{figure}[t!]
\epsfig{file=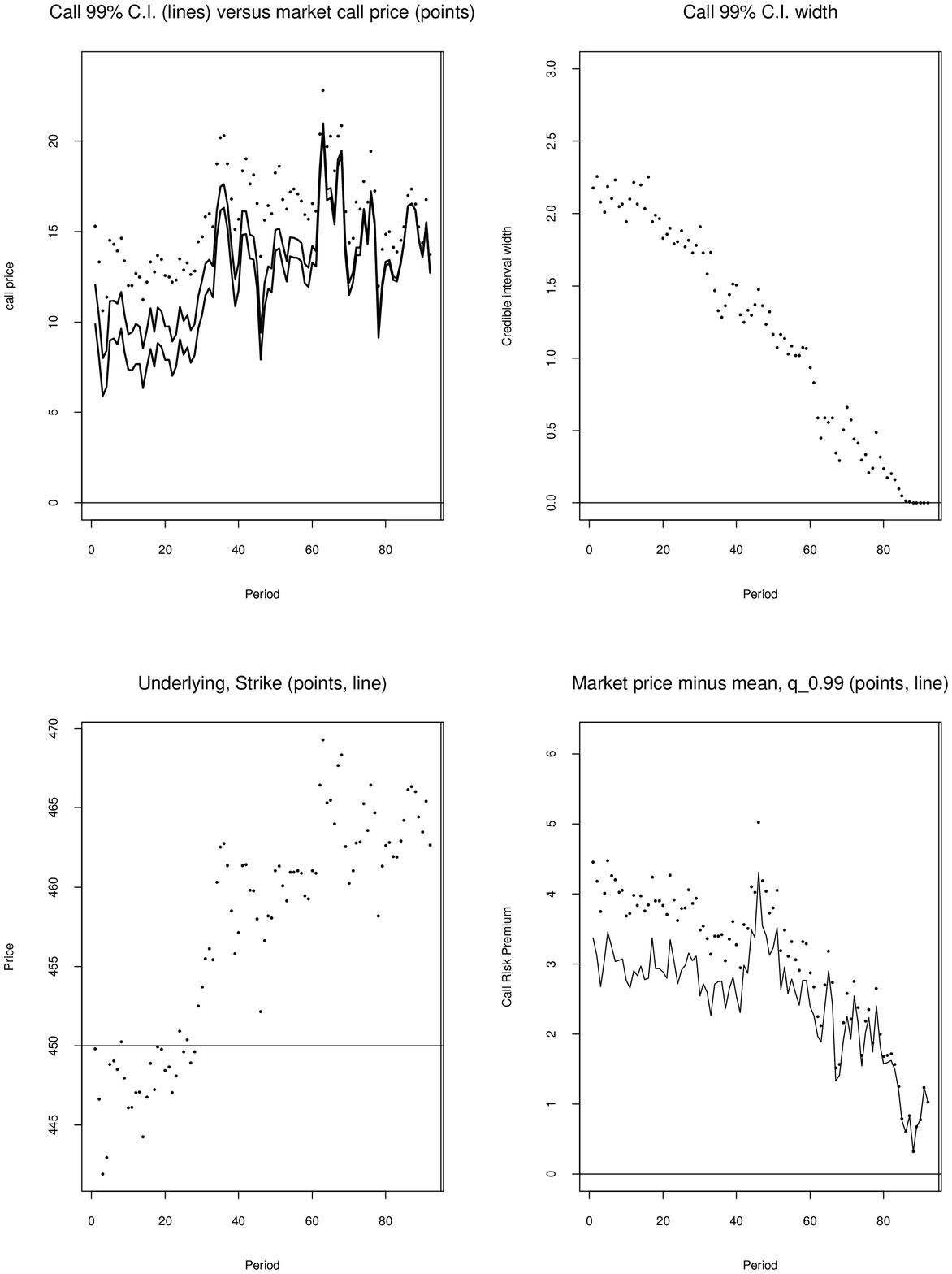,width=6in,height=4in}
\caption{99\% credible interval and market call prices (top-left),
99\% credible interval range (top-right), underlying level and
strike (bottom-left) and estimates of risk premium (bottom-right)
under the expected $\xi$ method} \label{fig:ALL_PLOTS_SP500_3}
\end{figure}
\end{center}

The top-left plot in figures \ref{fig:ALL_PLOTS_SP500_1},
\label{fig:ALL_PLOTS_SP500_2}, and \label{fig:ALL_PLOTS_SP500_3}
represent the call market prices (points) as we move closer to
maturity (vertical line). The x-axis represents time, and as we
move towards the right we get closer to the maturity date of that
call. The two lines represent the 0.5\% and 99.5\% percentiles for
the posterior distribution of the call prices, at each point in
time. We have run the MCMC analysis on a rolling basis for the
three methods based on the information available until each time
point, computing the Monte Carlo-based percentiles for each
posterior sample, and those lines represent the 99\% credible
intervals for the call prices for each period until maturity.
Several interesting things can be extracted from these
plots:\\

First, we can see that the range of the 99\% credible interval gets
narrower as we get closer to maturity. We can see this more clearly
in the top-right plot, where we see that range size over time. We
should expect to be more certain about where the call price should
be as we get closer to maturity, since any uncertainty we might have
about the inputs will have a smaller impact. This is more evident
for smooth payoffs, where small differences in the inputs only
become large differences in payoffs if there is a long time to
maturity. The overall extent of the input uncertainty will be a
function of time, so this
methodology is specially suitable for options with longer maturities.\\

Second, we can see the gap between the credible interval and the
actual market price. This gap, which represents the risk premium
for the call in the market, gets smaller as we get closer to
maturity. The markets are adding a larger nominal spread for
larger maturities, which is reasonable, since larger maturity
implies larger uncertainty and larger risks associated with the
instrument. We can see this more clearly in the bottom-right plot,
where we see the risk premium over time, expressed as market call
price minus expected call price under the posterior (points) or
market call price minus 99\% percentile under the posterior
(line). In all three cases, as we get closer to maturity, the risk
premium goes to
zero.\\

Third, we can see that whenever there is a jump in the underlying,
a strong movement in the market, the risk premium tends to
increase. The bottom-left plot represents the underlying (points)
versus the strike (line). For example, we can see that there is a
strong price movement on the 42nd point. The underlying falls
rapidly, and at the same time, we can see in the top left plot how
both the market call price and the credible interval move
accordingly. What is more interesting is the behavior of these
during this period. First, we can see that the credible interval
has a wider range at that point (top-left plot), showing a larger
uncertainty about the true call price. Second, we can see that the
actual market risk premium increases in figures
\ref{fig:ALL_PLOTS_SP500_1} and \ref{fig:ALL_PLOTS_SP500_3}
(bottom-right plot) showing that the market not only repriced the
call by shifting its value down, but it did it in such
a way that the actual premium increased.\\

Fourth, from figure \ref{fig:POSTERIOR_MODEL DATA}, we observe
that parameter uncertainty because of lack of data means higher
option prices.\\

We only need to run one MCMC analysis per instrument and time
period. The MCMC algorithm can be partitioned and parallelized, as
we mention in the appendix. However, the major advantage comes from
the Monte Carlo step, where we can indeed fully parallelize the
algorithm to compute the (iid) Monte Carlo samples. This is
especially useful if there were time constraints in the pricing
and/or the pricing algorithm was slow. In any case, the level of
precision required will be the determining factor of the actual
speed of the algorithm. In our example, it took a few seconds to run
each MCMC step, and the major computational cost came from the Monte
Carlo step for options with very long maturities, as it takes longer
to price the tree. Still the algorithm is quite tractable and simple
to code, and we feel the additional computational burden is very
limited compared to the additional information it provides. Of
course, for closed-form pricing models, these steps are even more
trivial and quick to construct.

\subsection{Comparison with na\"{i}ve calibration methods}\label{classicalcalibration}
This section includes a comparison between the three bayesian
parametric results from the mixture model proposed and several
possible naive calibration approaches that practitioners might
consider. We again use the S\&P500 data in a rolling fashion as we
did in the previous example, and will assume that there is no
options market.

The naive calibration procedures we consider are as follows:
\begin{itemize}
\item Sample Means ($SM$): For each (rolling) sample of returns that
we use for calibration of the models, we take the sample mean of the
returns larger/smaller than the libor, which will be our up/down
moves. This provides a point estimate of the theoretical value of
the option, without a confidence region to account for errors around
it.
\item Bootstrapped means ($BM$): For each (rolling) sample of returns
that we use for calibration of the models, we take random samples
(with replacement) of the observed data, with sample lengths equal
to the number of up/down returns observed in the original (rolling)
sample. For each sample we compute the up/down means. We take 5000
such pairs of means and compute the corresponding call prices (on a
rolling basis). This provides us with (rolling) confidence regions.
\item Bootstrapped values ($BV$): For each (rolling) sample of returns
that we use for calibration of the models, we take random samples of
length 1 of the up and down returns. We take 5000 such pairs of
random data points and compute the corresponding call prices (on a
rolling basis). This provides us with (rolling) confidence regions.
\end{itemize}

\begin{center}
\begin{figure}[t!]
\epsfig{file=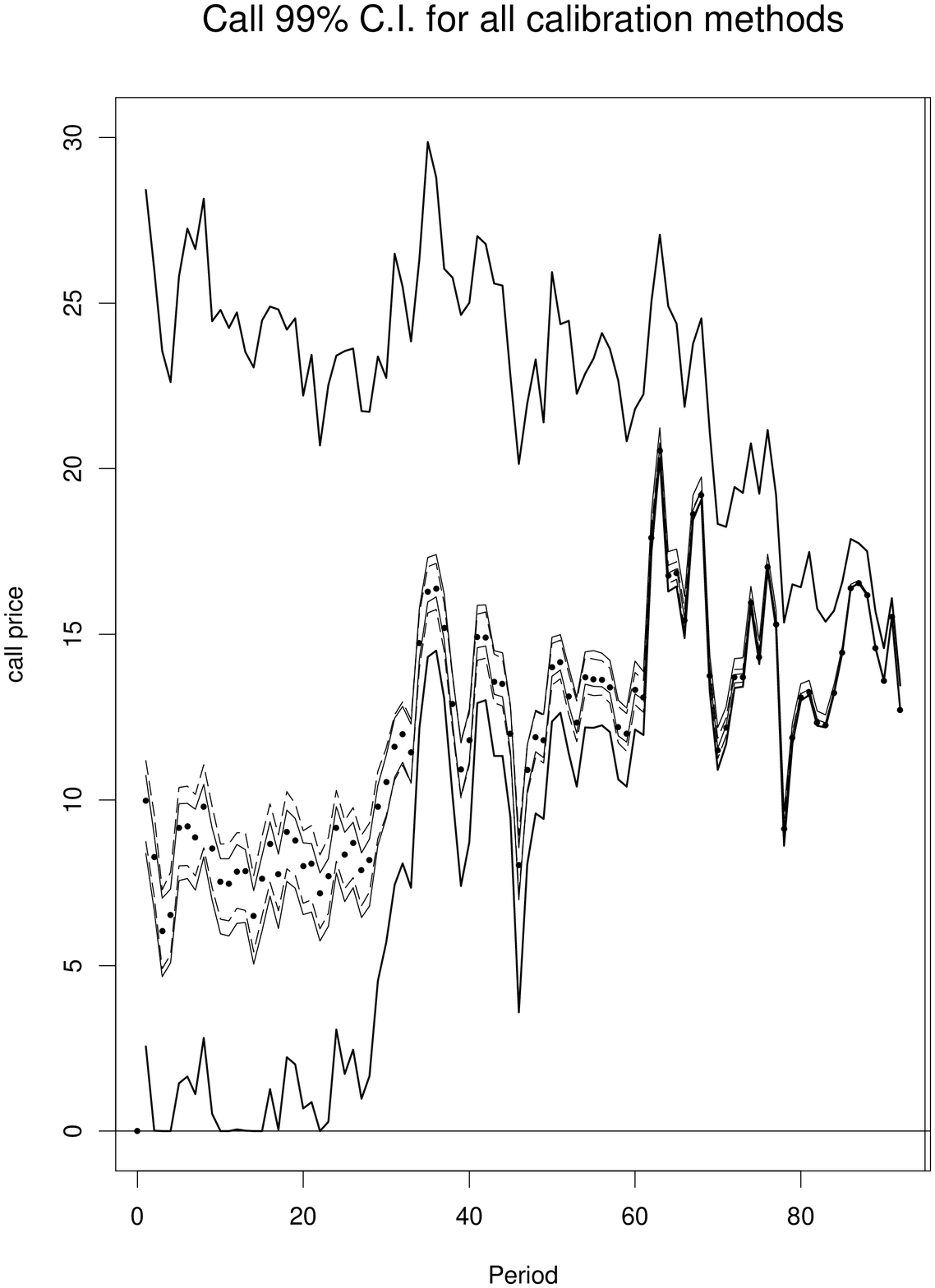,width=6in,height=4in}
\caption{99\% intervals for the Bootstrapped values approach
(thick line), bayesian approach (thin line), bootstrapped means
(dashed) and the sample means calibrated estimates (points) under
the $\theta$ method.} \label{fig:COMPARISON_1}
\end{figure}
\end{center}

\begin{center}
\begin{figure}[t!]
\epsfig{file=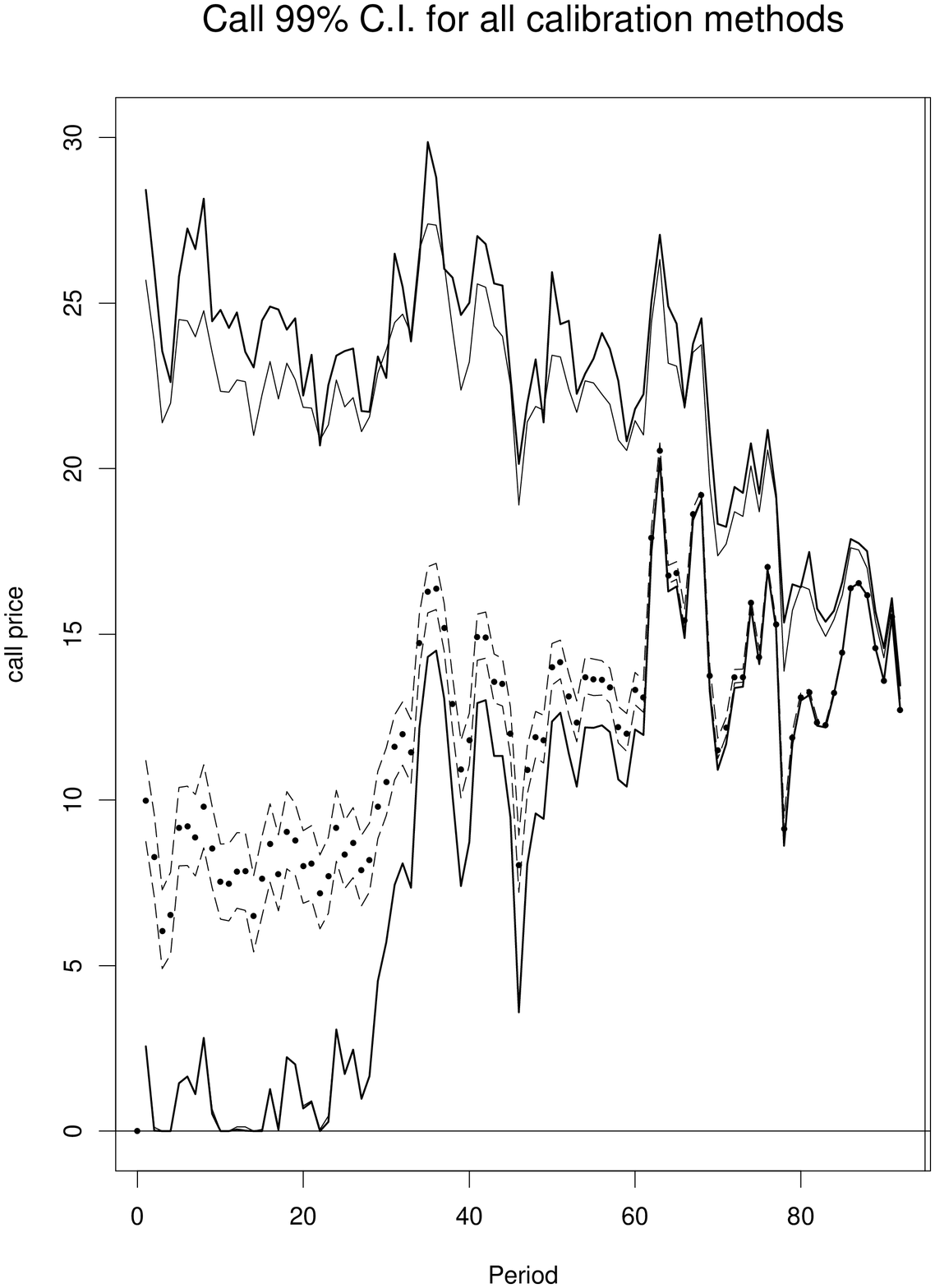,width=6in,height=4in}
\caption{99\% intervals for the Bootstrapped values approach
(thick line), bayesian approach (thin line), bootstrapped means
(dashed) and the sample means calibrated estimates (points) under
the $\xi$ method.} \label{fig:COMPARISON_2}
\end{figure}
\end{center}

\begin{center}
\begin{figure}[t!]
\epsfig{file=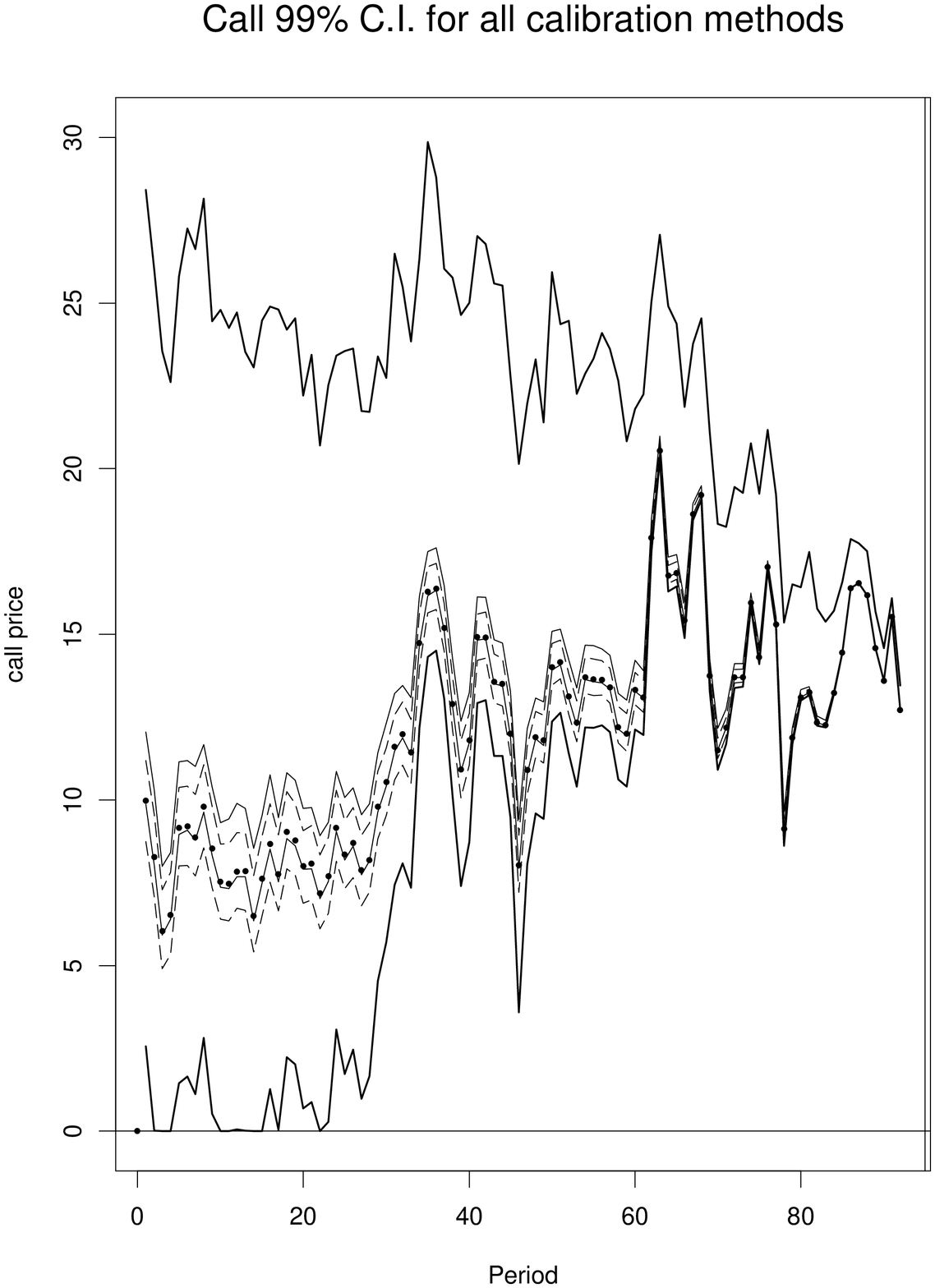,width=6in,height=4in}
\caption{99\% intervals for the Bootstrapped values approach
(thick line), bayesian approach (thin line), bootstrapped means
(dashed) and the sample means calibrated estimates (points) under
the expected $\xi$ method.} \label{fig:COMPARISON_3}
\end{figure}
\end{center}

Plots \ref{fig:COMPARISON_1}, \ref{fig:COMPARISON_2}, and
\ref{fig:COMPARISON_3} contain the 99\% bayesian credible interval
as well as each of the bootstrapped equivalent confidence
intervals. We perform the analysis in a similar rolling fashion
(with 252 business days of rolling
data) as we did in the previous subsection.\\
The points represent the calibrated values under the $SM$ method.
Notice that those values will account for variability in the final
payoff (through the trees), but not for parameter uncertainty. We
show in Table 1 that these point estimates are very close to those
under the bayesian parametric approach. Indeed, in terms of
interval width and mean values, the $\xi$ method and the BV are
similar. The same similarity applies to the $\theta$ and the expected $\xi$
methods compared to the BM method.\\
The dotted line represents the 99\% confidence interval based on the
$BM$ method. The confidence intervals are clearly very narrow.\\
The thin continuous line represents the 99\% bayesian credible
interval, while the thick line represents the equivalent under the
$BV$ method. We can see that the results are in this case quite
close. In any case all methods converge as we approach maturity, and
the option theoretical value is more certain.\\


\begin{center}
\begin{figure}[t!]
\epsfig{file=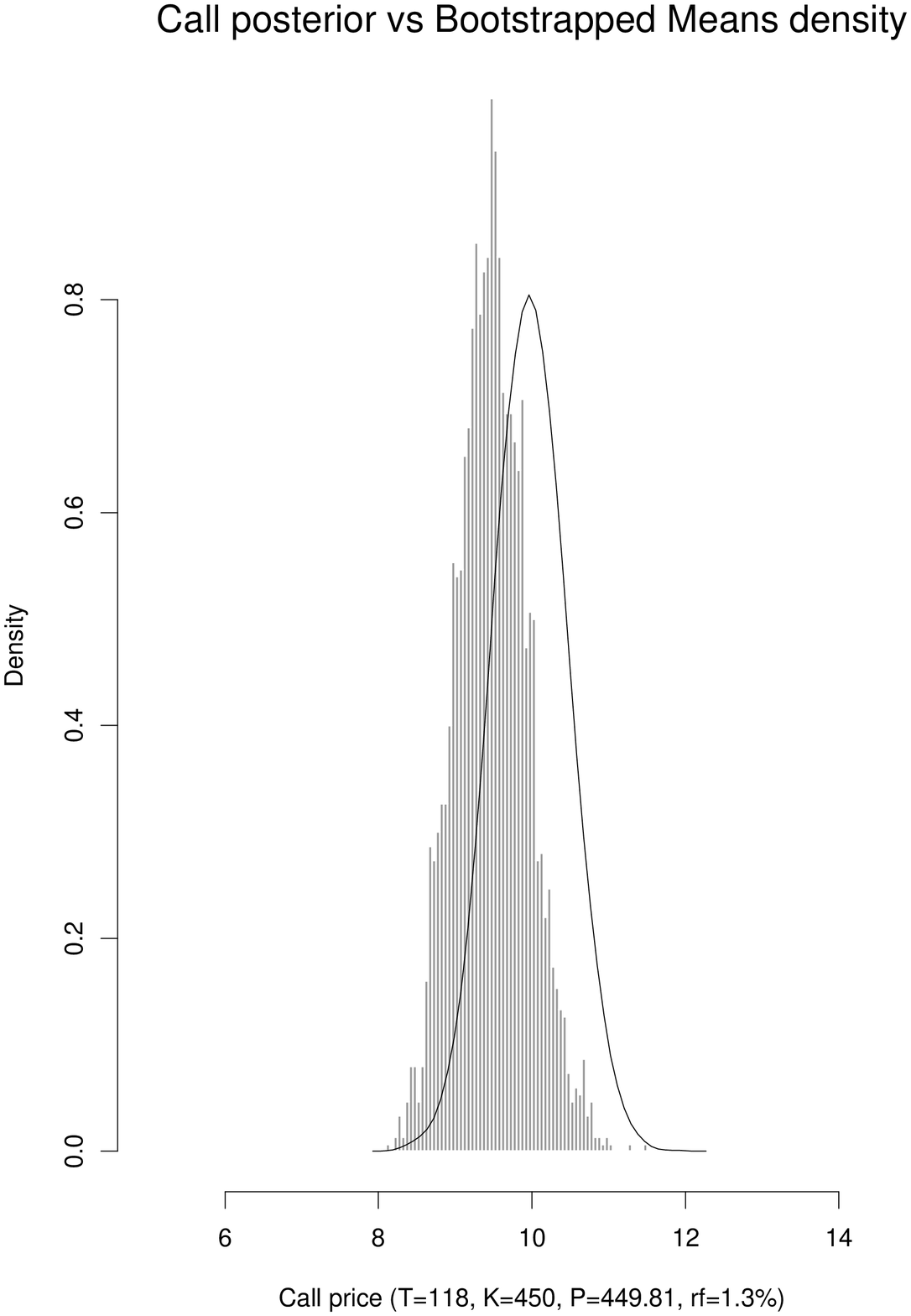,width=6in,height=4in}
\caption{Posterior distribution for a call price (histogram)
versus the density estimate based on bootstrapping means from the
sample ($BM$ calibration), both based on 5,000 Monte Carlo
samples. $\theta$ method.} \label{fig:PVB1}
\end{figure}
\end{center}

\begin{center}
\begin{figure}[t!]
\epsfig{file=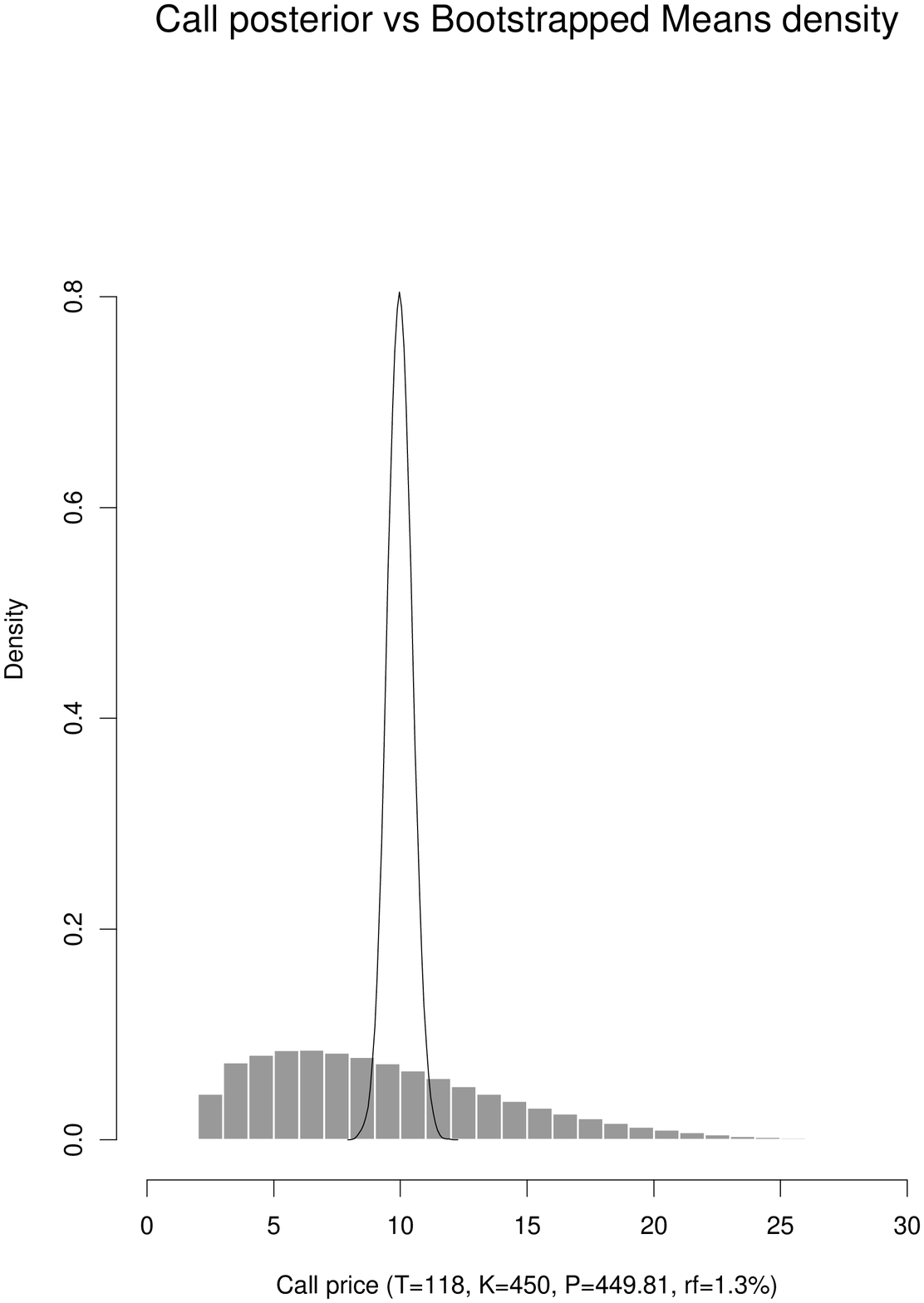,width=6in,height=4in}
\caption{Posterior distribution for a call price (histogram)
versus the density estimate based on bootstrapping means from the
sample ($BM$ calibration), both based on 5,000 Monte Carlo
samples. $\xi$ method.} \label{fig:PVB2}
\end{figure}
\end{center}

\begin{center}
\begin{figure}[t!]
\epsfig{file=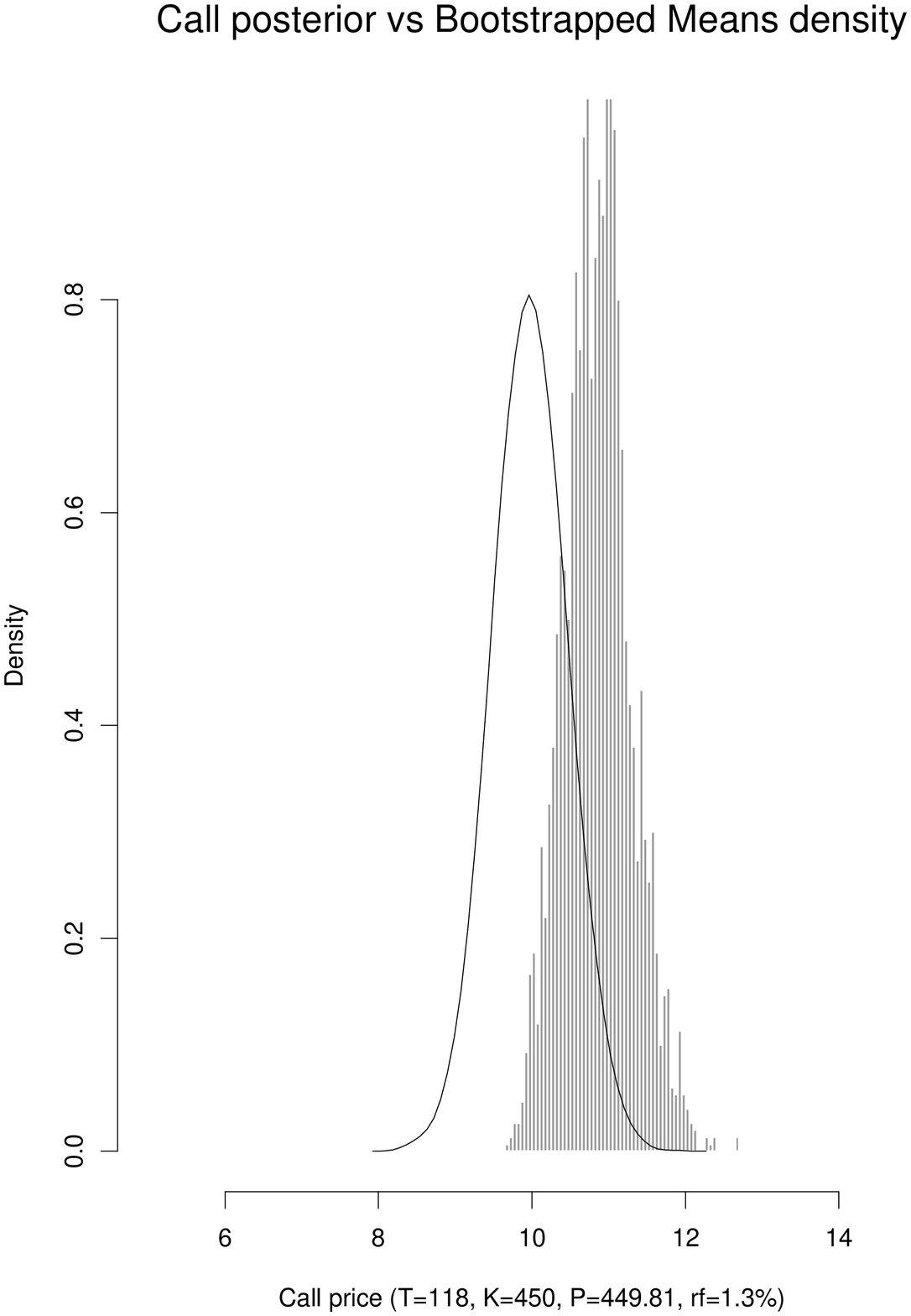,width=6in,height=4in}
\caption{Posterior distribution for a call price (histogram)
versus the density estimate based on bootstrapping means from the
sample ($BM$ calibration), both based on 5,000 Monte Carlo
samples. Expected $\xi$ method.} \label{fig:PVB3}
\end{figure}
\end{center}

\begin{table}[h!]\label{table}
\caption{Summary of results for the S\&P500 for different
calibration methods}
\begin{tabular}{lrrrrrr}
\hline\hline & $\xi$ Method & $\theta$ Method & Expected $\xi$
Method & SM & BM & BV
\\\hline
$Mean$                  & 12.04 & 12.52 & 12.04 & 12.06 & 12.06 & 11.54\\
$Median$                & 12.46 & 12.62 & 12.46 & 12.22 & 12.27 & 12.30\\
$99\%$ Range width      & 12.49 & 1.17  & 1.25  & 0.00  & 1.30 &
13.57
\\\hline
\end{tabular}\\
\noindent{}Means, medians, and 99\% range for the call prices on
the S\&P500 data. All values are averages over Monte Carlo samples
and over maturities.
\end{table}

There are several reasons that explain why we should use the
Bayesian approach in this paper instead of using the bootstrapping
one:
\begin{itemize}
\item A bootstrapping approach would suffer under low sample
sizes. For example, under a 1-period maturity and a sample of $n$
points, there are at most $n$ possible call prices. As $n$ gets
smaller, this would hinder the ability to create bootstrapped
intervals. The estimation of the tail of the call price distribution
would be quite unreliable. For example, the 1\% and 5\% quantiles
for the call price would be equal under a sample size of $n=20$.
\item A bootstrapping approach would suffer from outliers under low
sample sizes more than a parametric approach, that would be more
flexible to model/identify them. For example, we proposed a gaussian
mixture, but non-gaussian approaches would be possible, like
mixtures of t-distributions, that would account for outliers. A
small sample size with 1 outlier could heavily bias confidence
intervals under the bootstrapping approach. A parametric approach
allows to incorporate and impose the expected shape of the
population, as opposed to the sample shape. \item If we have a large
enough iid sample, the bootstrapping approach and our approach
provide quite similar results, as long as we correctly represent the
population. \item If the data is not iid, we can have more flexible
parametric approaches (markov switching, stochastic volatility),
which could not possible under the bootstrapping method. \item The
use of a bayesian approach (parametric and/or nonparametric) is more
appealing when in presence of low sample sizes, and where prior
process or parameter knowledge can be incorporated into the
analysis. \item Simple variations would allow parameter learning,
which would be of special relevance on overly-trending markets. For
example, if we assumed a common variance in the up/down moves,
having a small sample of down moves would be a problem under the
bootstrap approach, whereas it would not be a problem under the
parametric setting, as it would learn from the distribution of the
positives about the variability of the negatives.
\end{itemize}
The use of a bayesian parametric approach to propagate uncertainty
is justified, therefore, not only by its ability to mimic benchmarks
in simple cases, but also adjust to situations when the
bootstrapping fails.

\section{Other potential applications}
\subsection{Application to instruments without an options market}
We have provided a method that utilizes the data available about
some instrument and linked that data to the parameters of the option
pricing formula through a statistical model. We then have proceeded
to estimate these parameters and pass the uncertainty about them,
through the pricing formula, into uncertainty about the outputs. It
is worth noting that at no point we needed to use implied or options
market prices to do this. Therefore, this method has a very natural
use in pricing of instruments for which there is not an options
market defined, and for which calibration-based methods that use
options prices fail to provide an answer (Rubinstein, M. (1994)).
This is the case for most real options (Mun, J. (2005)), as well as
client-specific options for which we might want to make a market or
anything else where limited data is available.

\subsection{Potential applications and extensions}
When managing a portfolio of options, it is of great importance to
compute and track the Value-at-Risk (VaR) which tells us that with
a certainty of $\alpha$ percent, we will not lose more than X
dollars in the next N days (Hull, J. 2006). Since derivatives are
non-linear instruments, one needs to map the option position into
an equivalent cash position in its underlying in order to then
proceed to compute the VaR. In our framework, the VaR is the
percentile of the


Option prices depend on the parameter $\xi$. Therefore, once $\xi$
is known one can compute the option price $P(\xi)$. Furthermore,
one can generate the price distribution of the option with one of
the three methods described above. $\alpha$ is the confidence
level and percentile of the profit and loss distribution of the
derivative that one can choose (usually equal to $0.95$). As we
notice, the risk measures such as the VaR will depend on
parameters (just as we saw with option theoretical values) that we
will have to estimate. We therefore need to integrate the VaR (and
any other coherent risk measure such as the Expected Shortfall)
analytically or numerically with respect to the posterior
distribution of $\pi(\xi|\text{data})$ given the data and current
information set. The same analysis applies to other coherent risk
measures. The VaR will be influenced by the uncertainty through
the posterior distribution of the model parameter vector $\xi$.
Furthermore, since the posterior distributions can be asymmetric
and skewed we get that
$E_{\xi|\text{data}}\left\{VaR(P(\xi))\right\} =
\int_{\Xi}{VaR\left\{P(\xi)\right\}\pi(\xi|\text{data})d\xi}\neq
VaR\left\{P(E_{\xi|\text{data}}(\xi))\right\}$.

When generating trees through the sampling of the posterior
distribution of the model parameter through any of the three
methods, we can produce the posterior distribution of the hedging
ratio\footnote{See Hull (2006).} $\delta$. This distribution,
together with the observed underlying returns, allows computation
of profit and loss distribution of the derivative. Also, it is
worth noticing that the option pricing statistical link that we
have developed throughout this paper can be expanded and enhanced.
Several improvements could include modelling the variance of the
returns of the ups and downs through two stochastic volatility
models (Jacquier et al., 1994), or modelling truncated t-students
instead of truncated normals.

\section{Conclusion}

The problem of finding theoretical option values is one where, as
we saw in section \ref{utilities}, is a non-linear function of
several inputs, making it therefore highly sensitive to small
variations from its inputs. We therefore illustrated why it is
more adequate to use the whole posterior distributions from model
parameter as inputs instead of their most likely values. We then
proceeded to show the effects of parameter uncertainty into model
outputs and how this should be considered as a joint problem when
defining option pricing tools. The link between the pricing tools
in mathematical finance and the inferential tools from modern
statistics should be stronger if we are to provide full and more
accurate answers not only about our current knowledge, but about
our ignorance as well.

In section \ref{foundations} we showed how to construct a
posterior distribution on the space of model trees for option
pricing indexed by $\xi$. A posteriori, we described three related
methods for model calibration and determination of posterior
option price probability distribution.

In section \ref{results} we commented why the bootstrap approach
suffers from low sample sizes, hindering the ability to create
bootstrapped intervals that would make the estimation of the tail
of the call price distribution quite unreliable. For a large
sample, the bootstrap method and our approach provided similar
results. However, the use of a bayesian approach (parametric
and/or nonparametric) is more appealing when in presence of low
sample sizes, and where prior process or parameter knowledge can
be incorporated into the analysis. Simple variations would allow
parameter learning, which would be of special relevance on
overly-trending markets, as well as computing theoretical option
values when the historical data on the underlying and option
prices are quite small.

As a concluding remark, the naive method of plugging into an
option pricing model the most likely value of the model parameters
poses the problem that the results might not be optimal in a
utility-based framework. Considering the whole probability
distribution of inputs to express uncertainty about outputs is one
of the advantages of our methodology, and allows for a full
bayesian update of the tree as we observe more and more
realizations of the underlying $\xi$. The drawback of our
methodology is computational, as the simulation needs are much
larger.


\newpage
\section{Appendix 1: Bayesian implementation of the MCMC sampler}

We derive in this appendix the full conditional posterior
distributions. For full details about Bayesian estimation methods
and algorithms see Chen et al. (2000) or Robert and Casella
(1999).\\

The likelihood $L(u^{\ast },\sigma _{u}^{2},d^{\ast },\sigma
_{d}^{2},p|\xi _{1},...,\xi _{N})$ times the priors is
proportional to:

\begin{eqnarray*}
&&\overset{N}{\underset{i=1}{\Pi }}\left[ pTN(\xi _{i}|u^{\ast
},\sigma
_{u},1+r_{f},+\infty )+(1-p)TN(\xi _{i}|d^{\ast },\sigma _{d},0,1+r_{f})%
\right] \times  \\
&&\times p^{a-1}(1-p)^{b-1}\times  \\
&&\times (\sigma _{u}^{2})^{-\alpha _{u}-1}\exp \left[ -\frac{\beta _{u}}{%
\sigma _{u}^{2}}\right] (\sigma _{d}^{2})^{-\alpha _{d}-1}\exp \left[ -\frac{%
\beta _{d}}{\sigma _{d}^{2}}\right] 1_{\left\{ 0<d^{\ast
}<1+r_{f}<u^{\ast }<2\right\} }
\end{eqnarray*}

Rewriting it in an easier form we get:

\bigskip
\begin{eqnarray*}
&\propto &p^{a-1+\sum_{i=1}^{N}1_{\left\{ \xi _{i}>1+r_{f}\right\}
}}(1-p)^{b-1+\sum_{i=1}^{N}1_{\left\{ 0<\xi _{i}<1+r_{f}\right\}
}}1_{\left\{ 0<d^{\ast }<1+r_{f}<u^{\ast }<2\right\} }\times  \\
&&\times (\sigma _{u}^{2})^{-\alpha _{u}-1}\exp \left[ -\frac{\beta _{u}}{%
\sigma _{u}^{2}}\right] (\sigma _{d}^{2})^{-\alpha _{d}-1}\exp \left[ -\frac{%
\beta _{d}}{\sigma _{d}^{2}}\right] \overset{N}{\underset{i:\xi _{i}>1+r_{f}}%
{\Pi }}\frac{\phi (\xi _{i}|u^{\ast },\sigma _{u})}{1-\Phi (\xi
_{i}|u^{\ast
},\sigma _{u},1+r_{f})}\times  \\
&&\times \overset{N}{\underset{i:0<\xi _{i}<1+r_{f}}{\Pi
}}\frac{\phi (\xi _{i}|d^{\ast },\sigma _{d})}{\Phi (\xi
_{i}|d^{\ast },\sigma _{d},1+r_{f})-\Phi (\xi _{i}|d^{\ast
},\sigma _{d},0)}
\end{eqnarray*}

Notice that the full conditional of $p$ does not depend on any other
parameter. Therefore we compute it in closed form. This eases the
computation significantly.

\subsection{Adaptive variance proposal over pre-burn-in period}
The proposals for the remaining parameters are simply
formulated as a random walk around the current value, with fixed
variances $v_u, v_d, v_{\sigma_u^2},v_{\sigma_d^2}$. To set these
variance we allow a pre-burn-in period. We monitor the acceptance
ratios for the Metropolis algorithm over this period, and adjust the
variance up or down to reach a target Metropolis acceptance ratio
between 10\% and 50\%. Then, after we reach this ratio for all
parameters, we fix
that variance and allow the MCMC to start in its regular form.\\

There are three major advantages of this kind of proposals. First,
it allows the user a more black-box approach, where the algorithm
will adjust itself to reach an acceptable mixing of the chain.
Therefore, it requires less user inputs to run, making it more
automatic and appealing for practitioners. Second, it makes most
of the actual metropolis ratios in the sampling algorithm simpler,
as the contributions of the proposals cancel on numerator and
denominator, due to the symmetry of the proposal
$p(\text{proposed}|\text{current})=p(\text{current}|\text{proposed})$.
Third, it actually does work quite effectively in practice,
requiring in our S\&P analysis less than 500 iterations in most
cases to achieve
good proposals.\\

The pseudo-code for the sampling procedure looks as follows:
\begin{enumerate}
\item Preset a number of iterations N, over which we are going to
monitor the acceptance frequencies of the Metropolis steps. We chose
N=100
\item Assign starting values for $v_u,v_d,\sigma_u^2,\sigma_d^2$. For
practical purposes we chose the variances of the positive (negative)
returns to set $v_u,\sigma_u^2$ ($v_d,\sigma_d^2$).
\item Assign starting values for
$u^\ast,d^\ast,\sigma_u^2,\sigma_d^2$. For practical purposes we
chose the means and variances of positive (negative) returns.
\item Set the iteration index n=1.
\item Sample $p$ from its full conditional.
\item Sample the remaining parameters according to the sampling algorithms detailed below using the current values for the proposal variances.
\item Record whether we accepted the proposal in the metropolis steps for each of the four parameters $u^\ast,d^\ast,\sigma_u^2,\sigma_d^2$.
\item If n$<$N, then set n=n+1 and go back to step 5.
\item If n=N iterations, then check the acceptance frequencies for $each$ parameter. If \subitem we accepted the proposal
more than N/2 times, reduce the proposal variance by half. \subitem
we accepted the proposal less than N/10 times, double the proposal
variance.\subitem we accepted the proposal between N/10 and N/2
times, keep the current proposal.
\item If we modified any proposal variance at step 9, then go back to 4. Otherwise, set the current values for the proposal variances and start the
actual MCMC.
\end{enumerate}

One could refine this even further, as we can partition the
parameter space (and the MCMC) into the up and down blocks, which
are independent, and parallelize the computations.

\subsection{Full conditionals and the MCMC sampler}

\noindent\textbf{Sampling p}\\\\

Draw $\left[ p|Data\right] \sim Beta\left[
a+\sum_{i=1}^{N}1_{\left\{ \xi _{i}>1+r_{f}\right\}
},b+\sum_{i=1}^{N}1_{\left\{ 0<\xi _{i}<1+r_{f}\right\} }\right]$

Notice that p is the proportion of ups versus downs (adjusted by the
prior proportion), and with variance tending to $0$ as $N$ tends to
$+\infty $. A sufficient statistics for this full conditional is the
number of up and down moves in the sample, defined with respect to
the risk-free rate.\\\\

\noindent\textbf{Sampling u}\\\\
Draw $u^{p}\sim N\left[ u^{p}|u^\ast,v_u\right]$ and set $u^{\ast
}=u^{p}$ with probability

\begin{equation*}
\min \left[ 1,\frac{1_{\left\{ 0<d^{\ast }<1+r_{f}<u^{p}<2\right\}
}\left[ \overset{N}{\underset{i:\xi _{i}>1+r_{f}}{\Pi }}\frac{\phi
(\xi _{i}|u^{p},\sigma _{u})}{1-\Phi (\xi _{i}|u^{p},\sigma
_{u},1+r_{f})}\right] }{1_{\left\{ 0<d^{\ast }<1+r_{f}<u^{\ast }<2\right\} }\left[ \overset%
{N}{\underset{i:\xi _{i}>1+r_{f}}{\Pi }}\frac{\phi (\xi _{i}|u^{\ast
},\sigma _{u})}{1-\Phi (\xi _{i}|u^{\ast },\sigma
_{u},1+r_{f})}\right] }\right]
\end{equation*}

\noindent\textbf{Sampling $d$}\\\\
Draw $d^{p}\sim N\left[ d^{p}|d^\ast, v_d \right]$ and set $d^{\ast
}=d^{p}$ with probability:

\begin{equation*}
\min \left[ 1,\frac{1_{\left\{ 0<d^{p}<1+r_{f}\right\} }\left[ \overset{N}{%
\underset{i:0<\xi _{i}<1+r_{f}}{\Pi }}\frac{\phi (\xi _{i}|d^{p},\sigma _{d})%
}{\Phi (\xi _{i}|d^{p},\sigma _{d},1+r_{f})-\Phi (\xi
_{i}|d^{p},\sigma _{d},0)}\right]}{1_{\left\{ 0<d^{\ast }<1+r_{f}\right\} }%
\overset{N}{\underset{i:0<\xi _{i}<1+r_{f}}{\Pi }}\frac{\phi (\xi
_{i}|d^{\ast },\sigma _{d})}{\Phi (\xi _{i}|d^{\ast },\sigma
_{d},1+r_{f})-\Phi (\xi _{i}|d^{\ast },\sigma _{d},0)} }\right]
\end{equation*}

\noindent\textbf{Sampling $\sigma_u^2$}\\\\
Draw $\sigma _{u}^{2,p}\sim N\left[ \sigma _{u}^{2,p}| \sigma_u^2,
v_{\sigma_u^2}\right]$ and set $\sigma _{u}^{2}=\sigma _{u}^{2,p}$
with probability:

\begin{equation*}
\min \left[ 1,\frac{\left( \sigma _{u}^{2,p}\right) ^{-\alpha
_{u}-1}\exp
\left( -\frac{\beta _{u}}{\sigma _{u}^{2,p}}\right) \left[ \overset{N}{%
\underset{i:\xi _{i}>1+r_{f}}{\Pi }}\frac{\phi (\xi _{i}|u,\sigma _{u}^{p})}{%
1-\Phi (\xi _{i}|u,\sigma _{u}^{p},1+r_{f})}\right] }{%
\left( \sigma _{u}^{2}\right) ^{-\alpha _{u}-1}\exp \left( -\frac{\beta _{u}%
}{\sigma _{u}^{2}}\right) \left[ \overset{N}{\underset{i:\xi _{i}>1+r_{f}}{%
\Pi }}\frac{\phi (\xi _{i}|u,\sigma _{u}^{\ast })}{1-\Phi (\xi
_{i}|u,\sigma _{u}^{\ast },1+r_{f})}\right] }\right]
\end{equation*}

We could in principle adjust the proposal to be a truncated normal,
so that we ensure that we never sample outside the parameter space,
and adjust the metropolis ratios with the normalizing constants.
However, in practice, we did not draw a single value smaller than
zero, so this adjustment, although technically more correct, would
not make any difference in practice, as the mass of the points below
zero for the proposal was effectively zero (under most choices of
$v$).\\

\noindent\textbf{Sampling $\sigma_d^2$}\\\\
Draw $\sigma _{d}^{2,p}\sim N\left[
\sigma _{d}^{2,p}| \sigma_d^2, v_{\sigma_d^2}\right]$ and set
$\sigma _{d}^{2}=\sigma _{d}^{2,p}$ with probability:

\begin{equation*}
\min \left[ 1,\frac{\left( \sigma _{d}^{2,p}\right) ^{-\alpha
_{d}-1}\exp
\left( -\frac{\beta _{d}}{\sigma _{d}^{2,p}}\right) \left[ \overset{N}{%
\underset{i:0<\xi _{i}<1+r_{f}}{\Pi }}\frac{\phi (\xi _{i}|d^{\ast
},\sigma _{d}^{p})}{\Phi (\xi _{i}|d^{\ast },\sigma
_{d}^{p},1+r_{f})-\Phi (\xi _{i}|d^{\ast },\sigma
_{d}^{p},0)}\right] }{\left( \sigma _{d}^{2}\right) ^{-\alpha
_{d}-1}\exp \left( -\frac{\beta _{d}}{\sigma
_{d}^{2}}\right) \left[ \overset{N}{\underset{i:0<\xi _{i}<1+r_{f}}{\Pi }}%
\frac{\phi (\xi _{i}|d^{\ast },\sigma _{d})}{\Phi (\xi _{i}|d^{\ast
},\sigma _{d},1+r_{f})-\Phi (\xi _{i}|d^{\ast },\sigma
_{d},0)}\right]}\right]
\end{equation*}

\section{Appendix: Markov Chain Monte Carlo summary}
In this section we include a small summary for one of the Markov
Chains we have ran.\\

\begin{center}
\begin{figure}[t!]
\epsfig{file=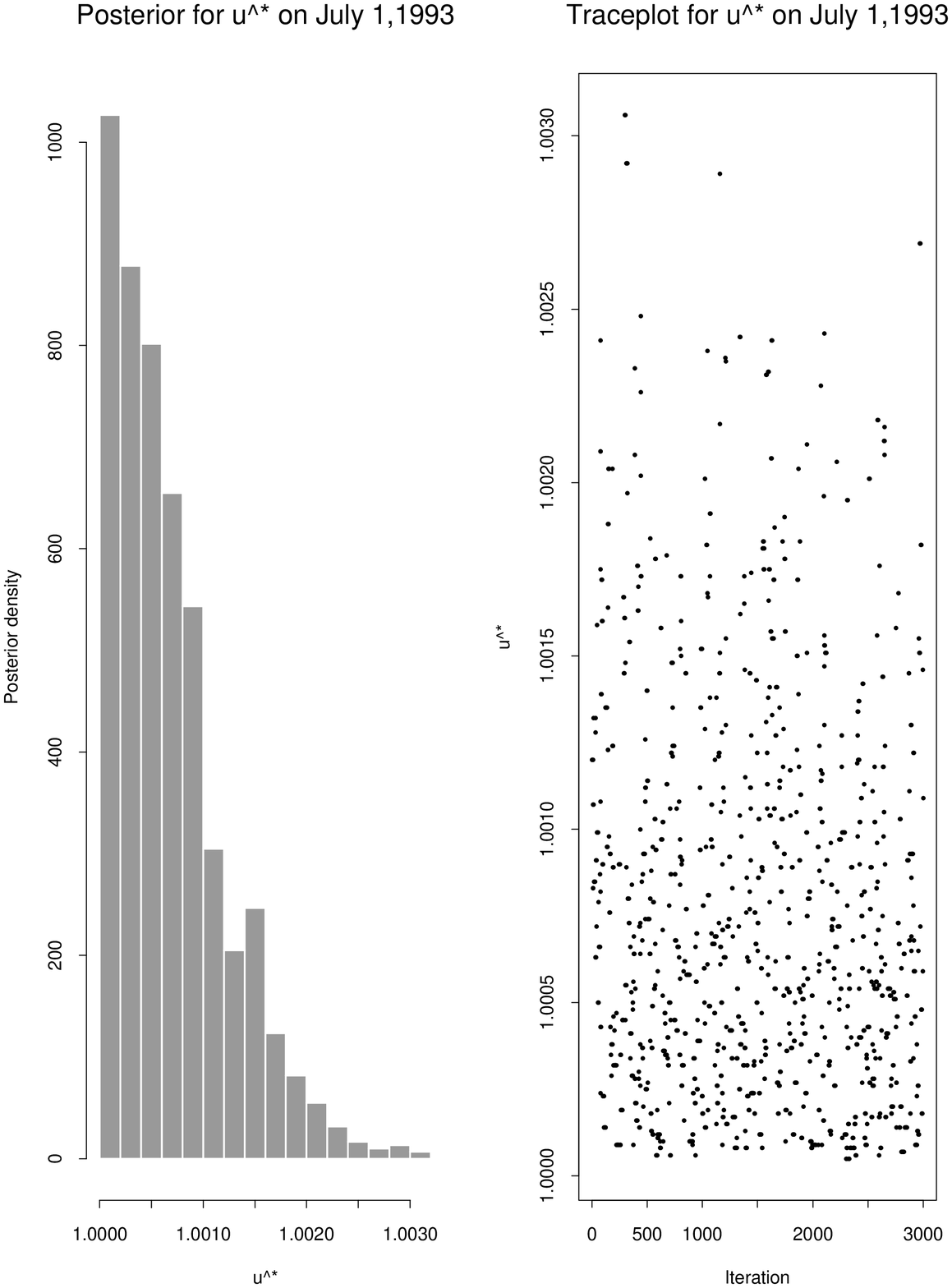,width=6in,height=4in}
\caption{Posterior distribution for the parameter $u^\ast$.}
\label{POSTERIOR_USTAR}
\end{figure}
\end{center}

\begin{center}
\begin{figure}[t!]
\epsfig{file=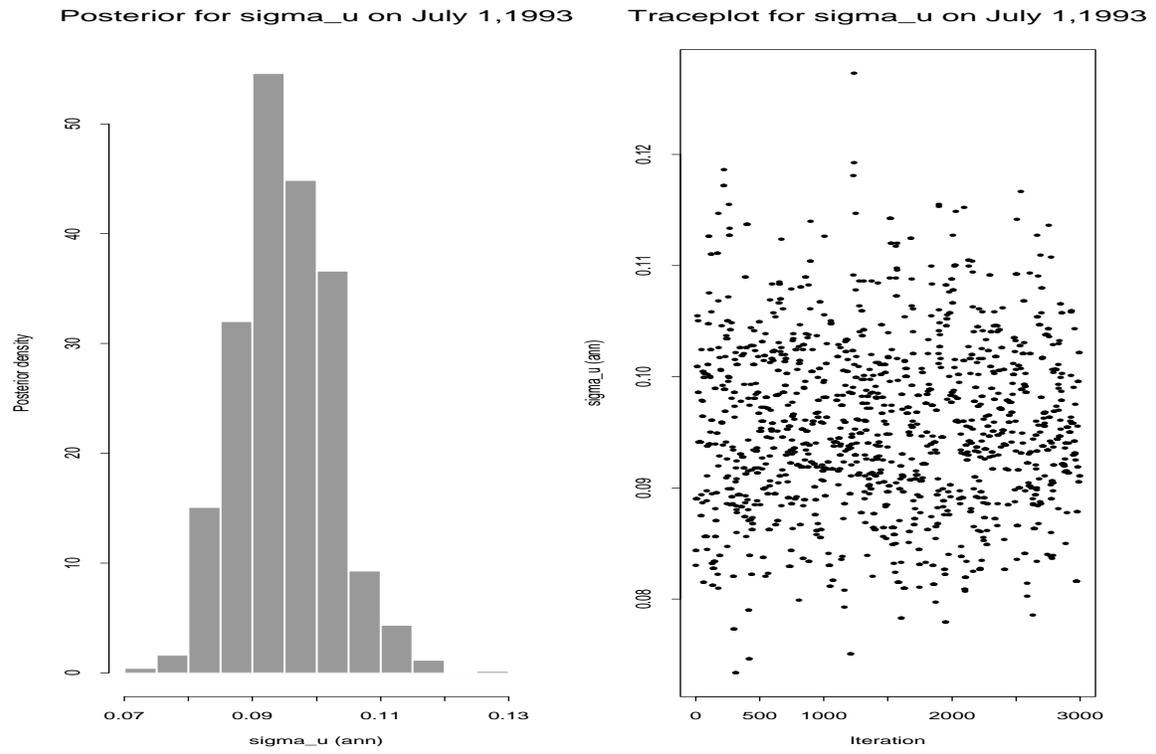,width=6in,height=4in}
\caption{Posterior distribution for the parameter $\sigma_{u}$.}
\label{POSTERIOR_SIGMA_U}
\end{figure}
\end{center}

Figure \ref{POSTERIOR_USTAR} shows the posterior distribution for
the parameter $u^\ast$. The left plot is a histogram of the
posterior distribution. We can see that it is truncated (at the
level of the risk-free rate) and skewed. The right plot shows the
traceplot of
the sampler, where we can see a good enough mixing of the chain.\\
Figure \ref{POSTERIOR_SIGMA_U} shows the posterior distribution for
the parameter $\sigma_u$. We can also see a histogram of the actual
posterior and the traceplot. The mixing also seems reasonable,
averaging around 40\% acceptances in the metropolis
ratios for each of the parameters.\\

\begin{center}
\begin{figure}[t!]
\epsfig{file=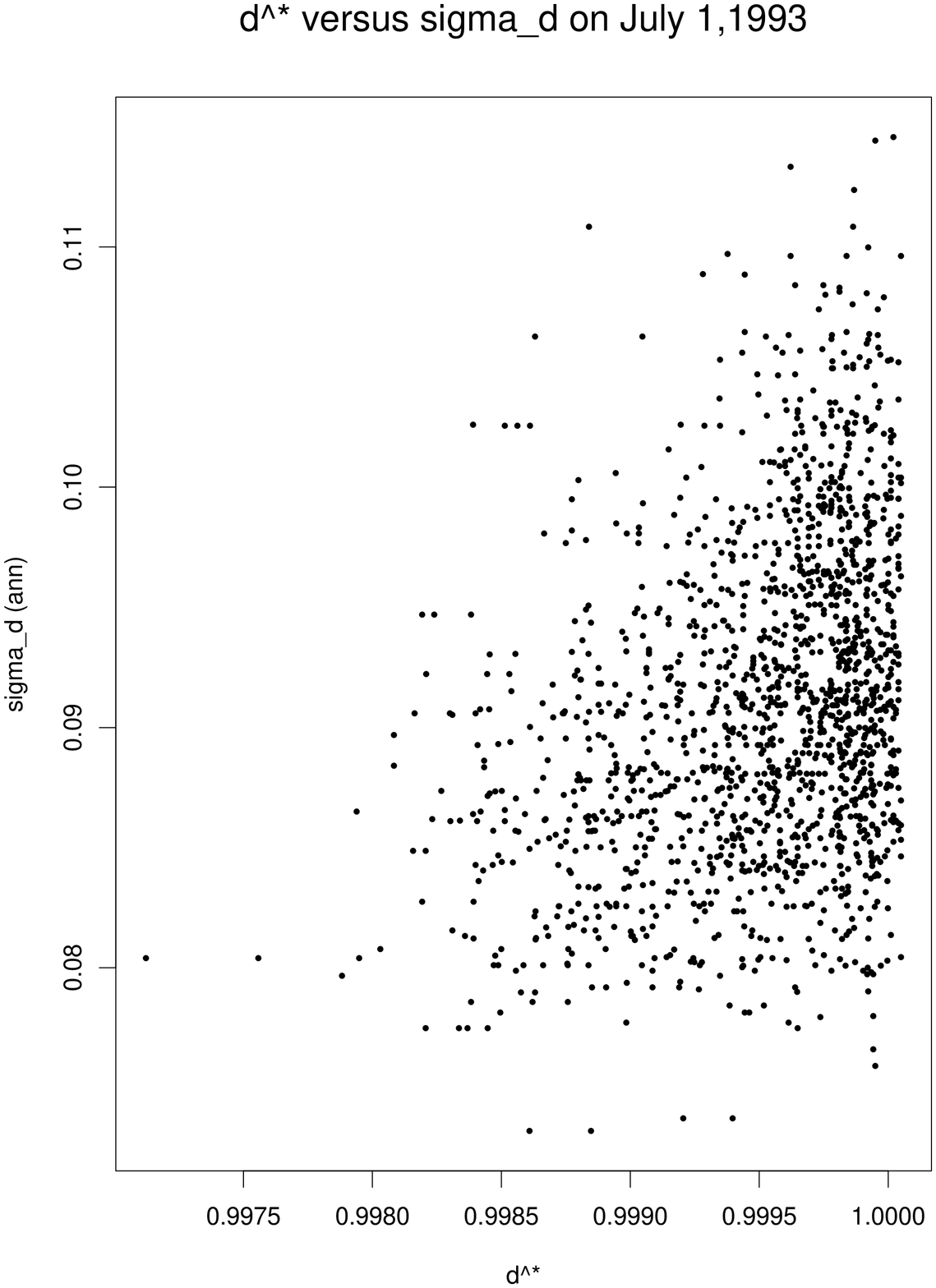,width=6in,height=4in} \caption{Joint
posterior distribution for the parameters $d^\ast$ and
$\sigma_d$.} \label{POSTERIOR_DSTAR_SIGMA_D}
\end{figure}
\end{center}

Since we can block the parameter space into three independent
blocks, given the structure of the joint distribution [$p$],
[$u^\ast$ and $\sigma_u$] and [$d^\ast$ and $\sigma_d$], we only
need to worry about the posterior correlation between parameters
within each block. Figure \ref{POSTERIOR_DSTAR_SIGMA_D} shows the
joint posterior distribution for $d^\ast$ and $\sigma_d$. We can see
that the posterior correlation is not excessive. Indeed, it averaged
35\% over the Markov Chains we ran (one per time until maturity of
the option).

\begin{center}
\begin{figure}[t!]
\epsfig{file=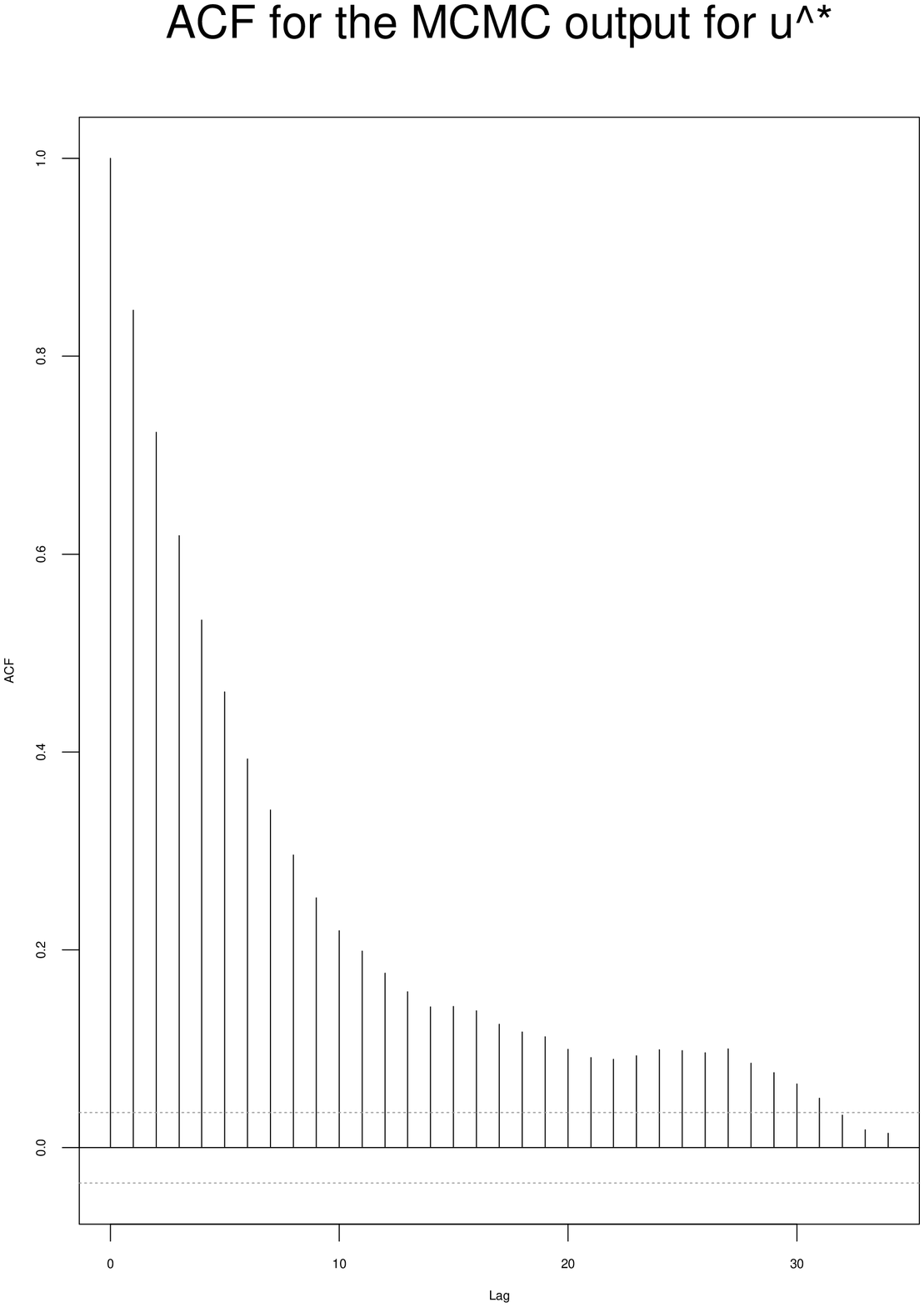,width=6in,height=4in}
\caption{Autocorrelation function of the MCMC output for one
parameter.} \label{ACF_USTAR}
\end{figure}
\end{center}

Finally we can see in figure \ref{ACF_USTAR} the autocorrelation
function for $u^\ast$ for one of the chains. It indicates us that
we can obtain pretty independent samples from the posterior with a
relatively small thinning of the chain.
\newpage

\end{document}